\documentclass[letterpaper,twocolumn,prl,aps,superscriptaddress,amsmath,amssymb,floatfix]{revtex4-1}
\usepackage{mathptmx}
\usepackage[utf8]{inputenc}
\usepackage{color}
\usepackage{amsmath}
\usepackage{amssymb}
\usepackage{graphicx}
\usepackage{esint}
\usepackage{enumitem}
\usepackage[unicode=true,
 bookmarks=true,bookmarksnumbered=false,bookmarksopen=false,
 breaklinks=false,pdfborder={0 0 1},backref=false,colorlinks=true]
 {hyperref}
\hypersetup{
 linkcolor=magenta,urlcolor=blue,citecolor=blue,pdfstartview={FitH}}

\makeatletter

\pdfpageheight\paperheight
\pdfpagewidth\paperwidth

\usepackage{textcomp}
\usepackage{epstopdf}

\pdfpageheight\paperheight
\pdfpagewidth\paperwidth

\@ifundefined{textcolor}{}{%
 \definecolor{BLACK}{gray}{0}
 \definecolor{WHITE}{gray}{1}
 \definecolor{RED}{rgb}{1,0,0}
 \definecolor{GREEN}{rgb}{0,1,0}
 \definecolor{BLUE}{rgb}{0,0,1}
 \definecolor{CYAN}{cmyk}{1,0,0,0}
 \definecolor{MAGENTA}{cmyk}{0,1,0,0}
 \definecolor{YELLOW}{cmyk}{0,0,1,0}
}

\usepackage{xcolor}
\usepackage{soul}
\definecolor{blue}{rgb}{0,0,1}
\definecolor{red}{rgb}{1,0,0}
\definecolor{green}{rgb}{0,1,0}

\usepackage{soul}

\makeatother

\begin{document}

\affiliation{Laboratory of Quantum Information, University of Science and Technology of China, Hefei 230026, China}
\affiliation{School of Information Science and Technology, ShanghaiTech University, Shanghai 201210, China}
\affiliation{Center for Quantum Information, Institute for Interdisciplinary Information Sciences, Tsinghua University, Beijing 100084, China}
\affiliation{Anhui Province Key Laboratory of Quantum Network, University of Science and Technology of China, Hefei 230026, China}
\affiliation{CAS Center For Excellence in Quantum Information and Quantum Physics, University of Science and Technology of China, Hefei, Anhui 230026, China}
\affiliation{Hefei National Laboratory, Hefei 230088, China}

\title{Acoustically control of integrated optical microrings: from photonic molecule to Mobius strip}

\author{Zheng-Xu~Zhu}
\thanks{These authors contributed equally to this work.}
\affiliation{Laboratory of Quantum Information, University of Science and Technology of China, Hefei 230026, China}

\author{Yuan-Hao~Yang}
\thanks{These authors contributed equally to this work.}
\affiliation{Laboratory of Quantum Information, University of Science and Technology of China, Hefei 230026, China}

\author{Xin-Biao~Xu}
\affiliation{Laboratory of Quantum Information, University of Science and Technology of China, Hefei 230026, China}

\author{Jia-Qi~Wang}
\email{wang1221@ustc.edu.cn}
\affiliation{Laboratory of Quantum Information, University of Science and Technology of China, Hefei 230026, China}

\author{Yu~Zeng}
\affiliation{Laboratory of Quantum Information, University of Science and Technology of China, Hefei 230026, China}

\author{Jia-Hua~Zou}
\affiliation{Laboratory of Quantum Information, University of Science and Technology of China, Hefei 230026, China}

\author{Juanjuan~Lu}
\affiliation{School of Information Science and Technology, ShanghaiTech University, Shanghai 201210, China}

\author{Weiting~Wang}
\affiliation{Center for Quantum Information, Institute for Interdisciplinary Information Sciences, Tsinghua University, Beijing 100084, China}

\author{Ming~Li}
\affiliation{Laboratory of Quantum Information, University of Science and Technology of China, Hefei 230026, China}
\affiliation{Hefei National Laboratory, Hefei 230088, China}

\author{Yan-Lei~Zhang}
\affiliation{Laboratory of Quantum Information, University of Science and Technology of China, Hefei 230026, China}
\affiliation{Hefei National Laboratory, Hefei 230088, China}

\author{Guang-Can~Guo}
\affiliation{Laboratory of Quantum Information, University of Science and
Technology of China, Hefei 230026, China}
\affiliation{Anhui Province Key Laboratory of Quantum Network, University of Science and Technology of China, Hefei 230026, China}
\affiliation{CAS Center For Excellence in Quantum Information and Quantum Physics,
University of Science and Technology of China, Hefei, Anhui 230026,
China}
\affiliation{Hefei National Laboratory, Hefei 230088, China}

\author{Luyan~Sun}
\email{luyansun@tsinghua.edu.cn}
\affiliation{Center for Quantum Information, Institute for Interdisciplinary Information Sciences, Tsinghua University, Beijing 100084, China}
\affiliation{Hefei National Laboratory, Hefei 230088, China}

\author{Chang-Ling~Zou}
\email{clzou321@ustc.edu.cn}
\affiliation{Laboratory of Quantum Information, University of Science and
Technology of China, Hefei 230026, China}
\affiliation{Anhui Province Key Laboratory of Quantum Network, University of Science and Technology of China, Hefei 230026, China}
\affiliation{CAS Center For Excellence in Quantum Information and Quantum Physics, University of Science and Technology of China, Hefei, Anhui 230026, China}
\affiliation{Hefei National Laboratory, Hefei 230088, China}

\date{\today}

\begin{abstract}
\textbf{Microring resonators (MRRs) are fundamental building blocks of photonic integrated circuits, yet their dynamic reconfiguration has been limited to tuning refractive index or absorption. Here, we demonstrate acoustic control over optical path topology on a lithium niobate on sapphire platform. By launching gigahertz acoustic waves into a hybrid phononic-photonic waveguide, a dynamic Bragg mirror (DBM) is created within the optical path, coupling forward and backward propagating light. Employing a pair of coupled MRRs, we achieve strong coupling between supermodes of the photonic molecule with only milliwatt-level drive power, yielding a cooperativity of 2.46 per milliwatt. At higher power, DBM reflectivity up to 24\% is achieved, revealing breakdowns of both the photonic molecule picture and perturbative coupled mode theory, indicating the transformation toward M\"{o}bius strip topology. Our work establishes a new dimension for controlling photonic devices, opening pathways toward fully reconfigurable photonic circuits through acoustic drive.}
\end{abstract}

\maketitle

\noindent \textbf{\large{}Introduction}{\large\par}
\noindent Since the 1990s, photonic integrated circuits (PICs) have emerged as indispensible platforms for modern optical information processing~\cite{bogaerts2020programmable,marpaung2019integrated,boes2023lithium,Shekhar2024,wabnitz2015all}, enabling complex functionalities within chip-scale footprints. More recently, PICs have also provided invaluable platforms for realizing scalable quantum circuits for quantum communication and computation~\cite{OBrien2009,Elshaari2020,Pelucchi2022,Cheng2023}. Among PIC building blocks, microring resonators (MRRs) are of fundamental importance due to their small mode volume and high quality factors, offering enhanced light-matter interaction~\cite{Liu2022} and versatile device functionality~\cite{bogaerts2012silicon,zhu2021integrated}. For example, MRRs have found applications in ultra-narrow linewidth lasers~\cite{jin2021hertz,liu2024fully}, wavelength-division multiplexing filters~\cite{liu2021silicon}, electro-optic modulators~\cite{xu2005micrometre,wang2018integrated,zhang2019broadband}, novel optical comb sources~\cite{Gaeta2019,Chang2022}, and biochemical sensors~\cite{ramachandran2008universal}. From a fundamental perspective, MRRs serve as testbeds for exploring nonlinear optics~\cite{strekalov2016nonlinear,eggleton2019brillouin,boes2023lithium} and quantum optics~\cite{Chang2018,Li2020,Zhou2023,Zhou2025} interactions. Their discrete spectral modes and tunable coupling make them also ideal for investigating emergent physics coupled MRRs, such as photonic topological insulators with protected edge states~\cite{ozawa2019topological,hafezi2013imaging}, exceptional points in non-Hermitian systems~\cite{hodaei2017enhanced}, and photonic molecules~\cite{zhang2019electronically} with controllable hybridization. 

\begin{figure*}[t]
\includegraphics[width=1\textwidth]{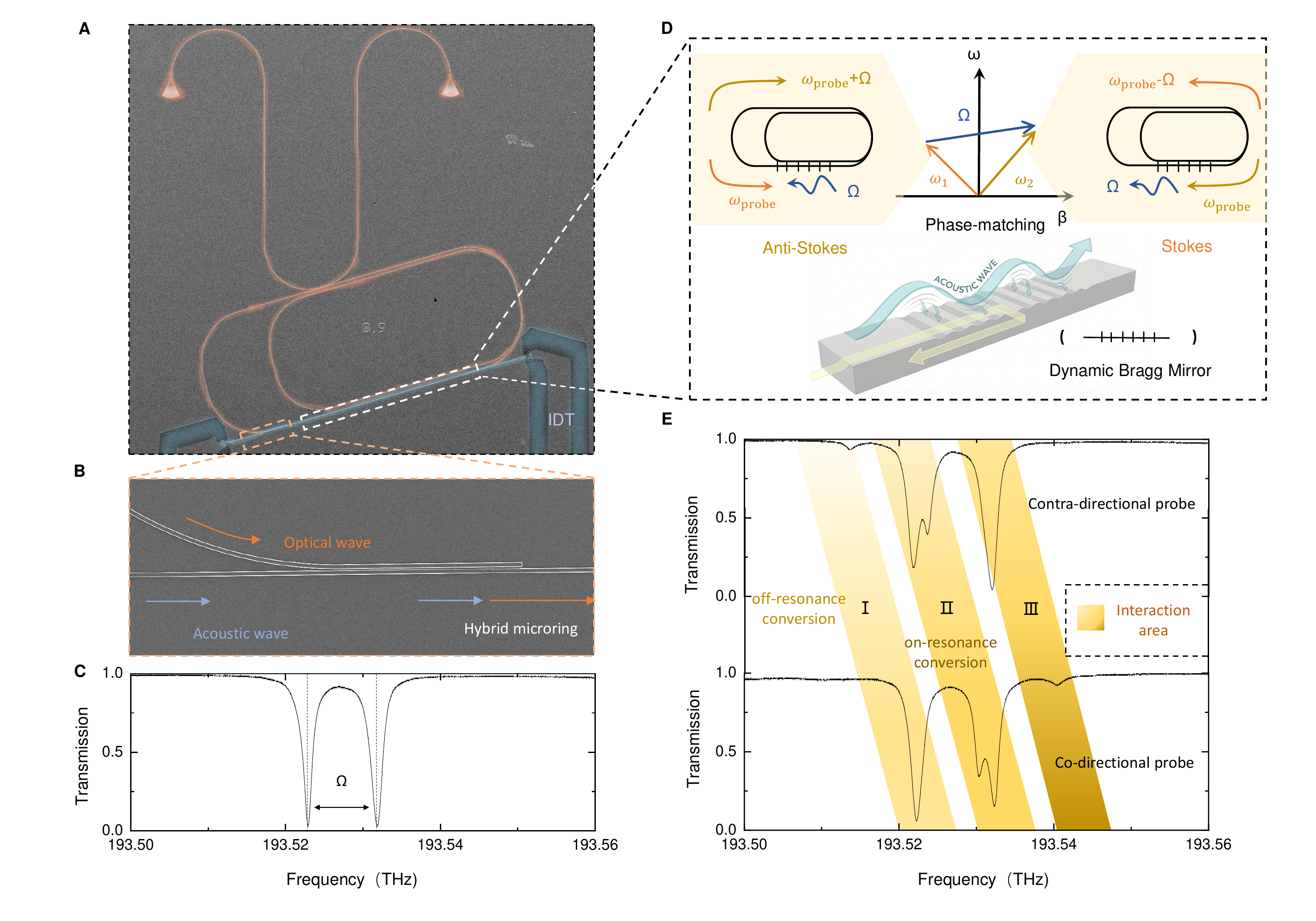}
\caption{\textbf{Acoustically driven dynamic Bragg mirror.} (A) The scanning electron microscope (SEM) image of the device. Optical components, including the gratings, waveguides, and hybrid microring resonator, are shown in orange. The interdigital transducers (IDTs) for acoustic wave generation are shown in blue. (B) The photon-phonon multiplexer. The optical mode (orange arrow) in the bending waveguide couples to the straight section of the hybrid microring resonator. The acoustic wave (blue arrow) is confined to and travels along this straight section, forming acousto-optic hybrid microring. (C) Measured optical transmission spectrum with the acoustic pump off.  (D) Schematic of the dynamic Bragg mirror (DBM) and phase-matching condition for the Brillouin interaction. An acoustic pump (blue lines, $\Omega$) mediates the coupling between two optical modes ($\omega_1$, orange and $\omega_2$, yellow). This non-reciprocal process causes a frequency up-conversion ($\omega_{\text{probe}} + \Omega$) or down-conversion ($\omega_{\text{probe}} - \Omega$) of the probe, depending on the relative propagation directions. (E) Experimental demonstration of strong coupling. With the acoustic pump around 0.6 mW activated, the transmission spectra are measured for a contra-directional probe (top panel, CCW) and a co-directional probe (bottom panel, CW). The spectra reveal clear phonon-induced mode splitting and frequency conversion. Highlighted areas distinguish the off-resonance conversion (I, III) and the on-resonance strong coupling (II) regimes.
}
\label{Fig1}
\end{figure*}

Dynamic reconfiguration of MRRs by actively controlling spectral properties, mode structure, and coupling topology is crucial for practical applications of MRRs. Various approaches have been developed to achieve such reconfiguration, including the electro-optic effect in lithium niobate~\cite{zhang2019electronically}, thermo-optic effects~\cite{vlasov2008high}, and magneto-optic effects~\cite{zhang2016optomagnonic,bi2011chip}. Absorption modulation can be achieved through electric field biasing of two-dimensional materials such as graphene~\cite{liu2011graphene,phare2015graphene}. Beyond material refractive index tuning of both real and imaginary components for phase and absorption control, device geometry can also be modified through electrostatic or optical gradient forces in suspended photonic structures~\cite{Poot2014,seok2016large}. Emerging non-volatile approaches have also been demonstrated for modulating MRRs, including phase-change materials that can be switched between amorphous and crystalline states~\cite{rios2015integrated} and optically-driven micro-actuators that modify refractive index distributions~\cite{Zhang2025}. However, these diverse methods are all limited to modifying the accumulated phase or attenuation of guided optical modes, limiting further exploration and control of optical path configurations.

Here, we demonstrate a fundamentally new dimension for controlling MRRs on a chip by acoustically driven backward Brillouin scattering. Through a piezoelectric transducer and RF input, high frequency traveling acoustic waves are excited to create dynamically moving Bragg gratings of refractive index modulation, effectively producing a dynamic Bragg mirror (DBM) that reflects guided light into reversed direction. The DBM reflectivity is electrically switchable, continuously tunable, and directionally reversible. Starting from a photonic molecule, the system exhibits remarkable strong coupling with increasing acoustic power, featuring a high normalized optical mode strong coupling cooperativity of $2.46\,\mathrm{mW}^{-1}$. A DBM reflectivity up to $24\%$ has been achieved, showing significant modification of optical paths in MRRs where both the photonic molecule picture and perturbative coupled mode theory break down. By providing electrical control over optical path topology rather than merely phase or absorption, our work opens pathways toward fully reconfigurable photonic circuits for dynamic signal routing, programmable optical filtering, and arbitrary dispersion control.

\smallskip{}
\noindent \textbf{\large{}Results}{\large\par}
\noindent Figure~\ref{Fig1}(A) shows a scanning electron micrograph (SEM) of our fabricated device on a thin-film lithium niobate on sapphire (LNOS) chip. This material platform is selected for realizing ``Zhengfu" architecture for scalable hybrid phononic and photonic integrated circuits~\cite{yang2024proposal,yang2025multi}, featuring simultaneous tight confinement of low-loss acoustic and optical modes without suspension, as well as efficient  interdigital transducers (IDTs) for converting RF signals to traveling acoustic waves. Backward Brillouin scattering provides the physical mechanism of our reconfigurable MRR: when an acoustic wave with frequency $\Omega$ is launched in the hybrid photonic-phononic waveguide, the photo-elastic effect creates a DBM through a grating of refractive index modulation, as illustrated in the inset of Fig.~\ref{Fig1}D. Consequently, optical photons can be reflected with frequency shifted by $\pm\Omega$ when the phase matching condition is satisfied [Fig.~\ref{Fig1}(D)].

The device consists of two evanescently coupled MRRs forming a ``photonic molecule", a pair of optical grating couplers for input/output, and two IDTs serving as active ports for bidirectional acoustic wave generation. Figure~\ref{Fig1}(B) provides a magnified view of the acoustic-input area, showing a specially designed photon-phonon multiplexer that maintain the acoustic wave propagating only in the straight waveguide section while allowing the optical wave to complete a round trip as an MRR. The two coupled MRRs have slightly different radii, creating a frequency splitting between their hybrid supermodes that is designed to match the acoustic wave frequency ({$\Omega/2\pi\sim 8.54\,\mathrm{GHz}$} for telecom optical wavelengths), enabling both the input and reflected frequency-shifted light to resonate within the coupled MRRs.

In general, the system can be described by the Hamiltonian
\begin{eqnarray}
    H & = & \omega_a (a_{\mathrm{cw}}^\dagger a_{\mathrm{cw}} + a_{\mathrm{ccw}}^\dagger a_{\mathrm{ccw}}) + \omega_b (b_{\mathrm{cw}}^\dagger b_{\mathrm{cw}} + b_{\mathrm{ccw}}^\dagger b_{\mathrm{ccw}}) \nonumber\\
    && + g( a_{\mathrm{cw}}^\dagger b_{\mathrm{cw}} +a_{\mathrm{cw}} b_{\mathrm{cw}}^\dagger + a_{\mathrm{ccw}}^\dagger b_{\mathrm{ccw}} +a_{\mathrm{ccw}} b_{\mathrm{ccw}}^\dagger) \nonumber\\
     && + \beta(a_{\mathrm{ccw}}^\dagger a_{\mathrm{cw}} e^{i\Omega t} +a_{\mathrm{ccw}} a_{\mathrm{cw}}^\dagger e^{-i\Omega t}).
    \label{Eq1}
\end{eqnarray}
Here, $a$ and $b$ are the bosonic annihilation operators for the optical modes outer and inner MRRs, respectively, with subscripts denoting clockwise (CW) and counter-clockwise (CCW) propagating modes, with resonant frequencies $\omega_{a}$ and $\omega_{b}$. $g$ is the coupling rate between the MRRs, and $\beta\propto\sqrt{P_\mathrm{a}}$ is the acoustically-induced backward Brillouin interaction strength determined by the acoustic drive power $P_\mathrm{a}$. Note that the model considers only the four coupled modes and neglects other modes separated by free spectral ranges (FSRs), valid  in the perturbative regime where the DBM reflectivity $R\ll1$, such that $\beta\approx R/\tau_{\mathrm{rt}}$, where $\tau_{\mathrm{rt}}$ is the optical round trip time in the outer MRR.

In the absence of an acoustic drive, the DBM has zero reflectivity $R=0$, and we obtain the optical transmission spectrum of the unperturbed system. As shown in Fig.~\ref{Fig1}(C), a pair of nearly identical Lorentzian dips reveals  two supermodes of the photonic molecule, with $\omega_a\approx\omega_b=\omega_0$ and $g \approx 2\pi \times 4.2 \text{ GHz}$ giving rise to a splitting of $2g\approx \Omega$. For simplicity, we can introduce symmetric supermodes $m_{\mathrm{cw,ccw}}=(a_{\mathrm{cw,ccw}}+b_{\mathrm{cw,ccw}})/\sqrt{2}$ and $n_{\mathrm{cw,ccw}}=(a_{\mathrm{cw,ccw}}-b_{\mathrm{cw,ccw}})/\sqrt{2}$. As illustrated in the phase-matching diagram in Fig.~\ref{Fig1}(D), the interaction is direction-dependent: for a CCW-propagating acoustic drive, an optical probe at frequency $\omega_{\text{probe}}$ that is contra-directional (CW) to the acoustic drive (CCW) is up-converted to a co-directional (CCW) mode at frequency $\omega_{\text{probe}} + \Omega$. Conversely, a probe that is co-directional (CCW) with the pump is down-converted to a contra-directional mode (CW) at $\omega_{\text{probe}} - \Omega$. When we activate the CCW DBM ($R\neq0$) by launching the acoustic drive, the two supermodes coupled, and the system Hamiltonian in the photonic molecule picture reads
\begin{equation}
    H / \hbar =
\begin{bmatrix}
\omega_0+g & 0 & \beta/2 & \beta/2 \\
0 & \omega_0-g & \beta/2 & \beta/2 \\
\beta/2 & \beta/2 & \omega_0+g-\Omega & 0 \\
\beta/2 & \beta/2 & 0 & \omega_0-g-\Omega
\end{bmatrix}
\label{Eq2}
\end{equation}
in the basis of four supermodes $\{ m_{\text{cw}}, n_{\text{cw}}, m_{\text{ccw}}, n_{\text{ccw}}\}$. Due to symmetry, the same derivation for a CW acoustic drive can be obtained by exchanging the subscripts cw and ccw.

Figure~\ref{Fig1}(E) demonstrates the reconfiguration of the MRRs through measured transmission spectra for both CW (top panel) and CCW (bottom panel) optical probes, under a fixed CCW acoustic drive power $P_\mathrm{a}=0.6\,\mathrm{mW}$. A distinct mode splitting is observed indicating the strong coupling between two optical modes~\cite{Guo2016}. The spectra exhibit two distinguishing features that showing the different effect of DBM from static mirrors. First, four dips arises because all four modes participate in the Brillouin interaction. Second, the CW and CCW probe spectra display anti-symmetry with respect to the symmetry axis of the unperturbed photonic molecule [Fig.~\ref{Fig1}(C)]. This anti-symmetry originates from the Hamiltonian symmetry under the exchange the CW and CCW mode with conjugation: probing in the CCW direction with reversed detuning yields the same spectrum as probing in the CW direction. Accordingly, we group the interactions into two categories as indicated by the shaded regions: regions I and III for off-resonance coupling, and region II for on-resonance coupling.

\begin{figure*}[t]
\includegraphics[width=1\textwidth]{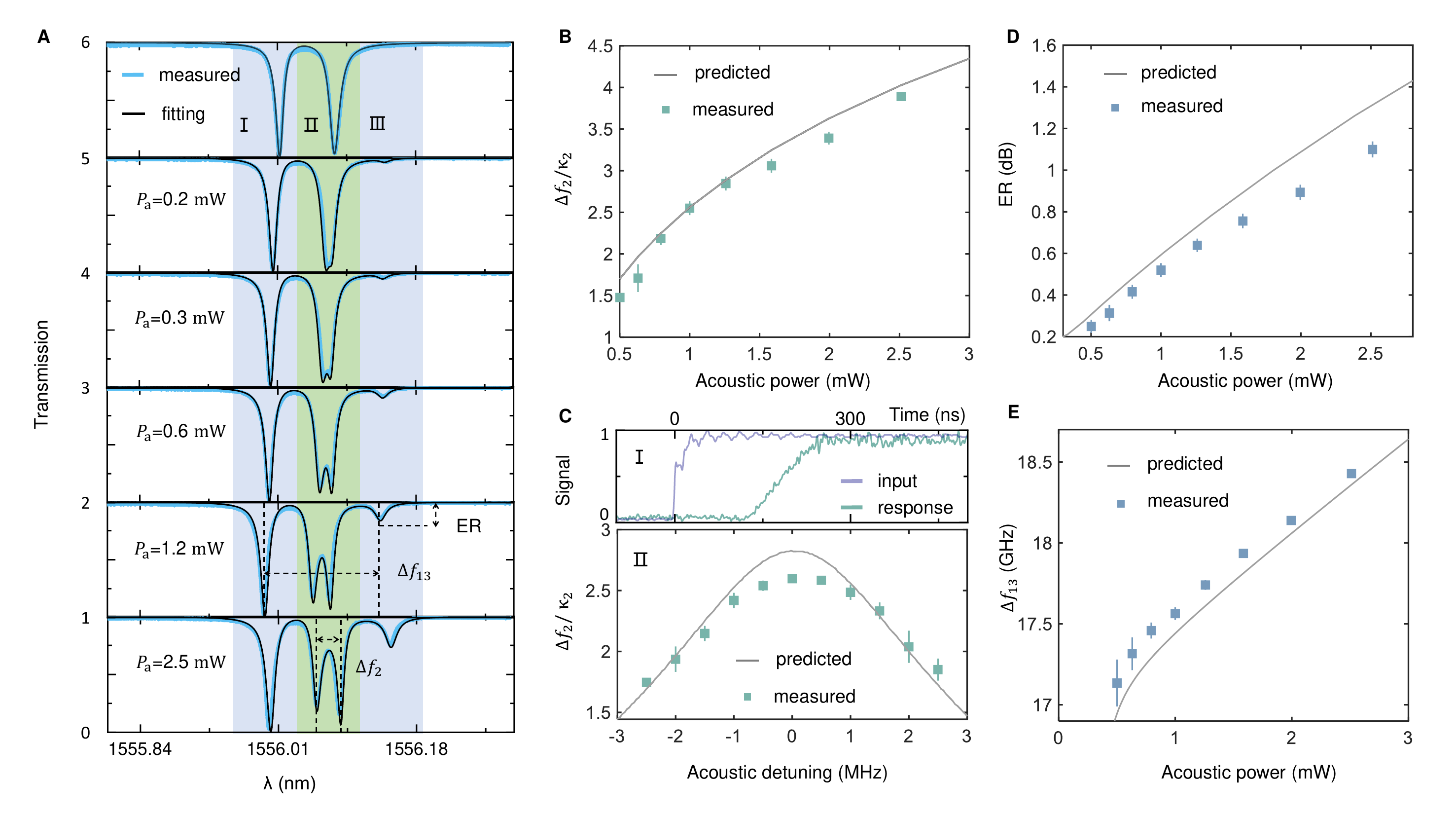}
\caption{\textbf{DBM induced reconfiguration of the microring and optical strong coupling.} (A) Measured transmission spectra (light blue lines) for increasing acoustic pump power, from top to bottom. Thin black lines represent the theoretical fits, which are used to extract key coupling parameters. (B) Normalized mode splitting, $\Delta f_2 / \kappa_2$ (where $\kappa_2= 0.90\,\text{GHz}$ is the linewidth), as a function of acoustic pump power. The data (blue squares) clearly follows the characteristic square-root dependence ($\propto \sqrt{P}$) predicted by theory (black line). (C) Normalized mode splitting $\Delta f_2 / \kappa_2$ versus acoustic detuning. (D) Extinction ratio (ER) of the newly generated mode (region III) versus acoustic pump power. (E) Relative frequency $\Delta f_{13}$ (between the reference mode I and the new mode III) versus acoustic pump power. Mode I is used as a reference to compensate for thermal drift. In (B-E), squares represent the data derived from the experimental fits in (A), while the dashed lines are theoretical predictions calculated using the parameters extracted from these fits. Error bars represent fitting uncertainties.}
\label{Fig2}
\end{figure*}

\begin{figure*}[t]
\includegraphics[width=1\textwidth]{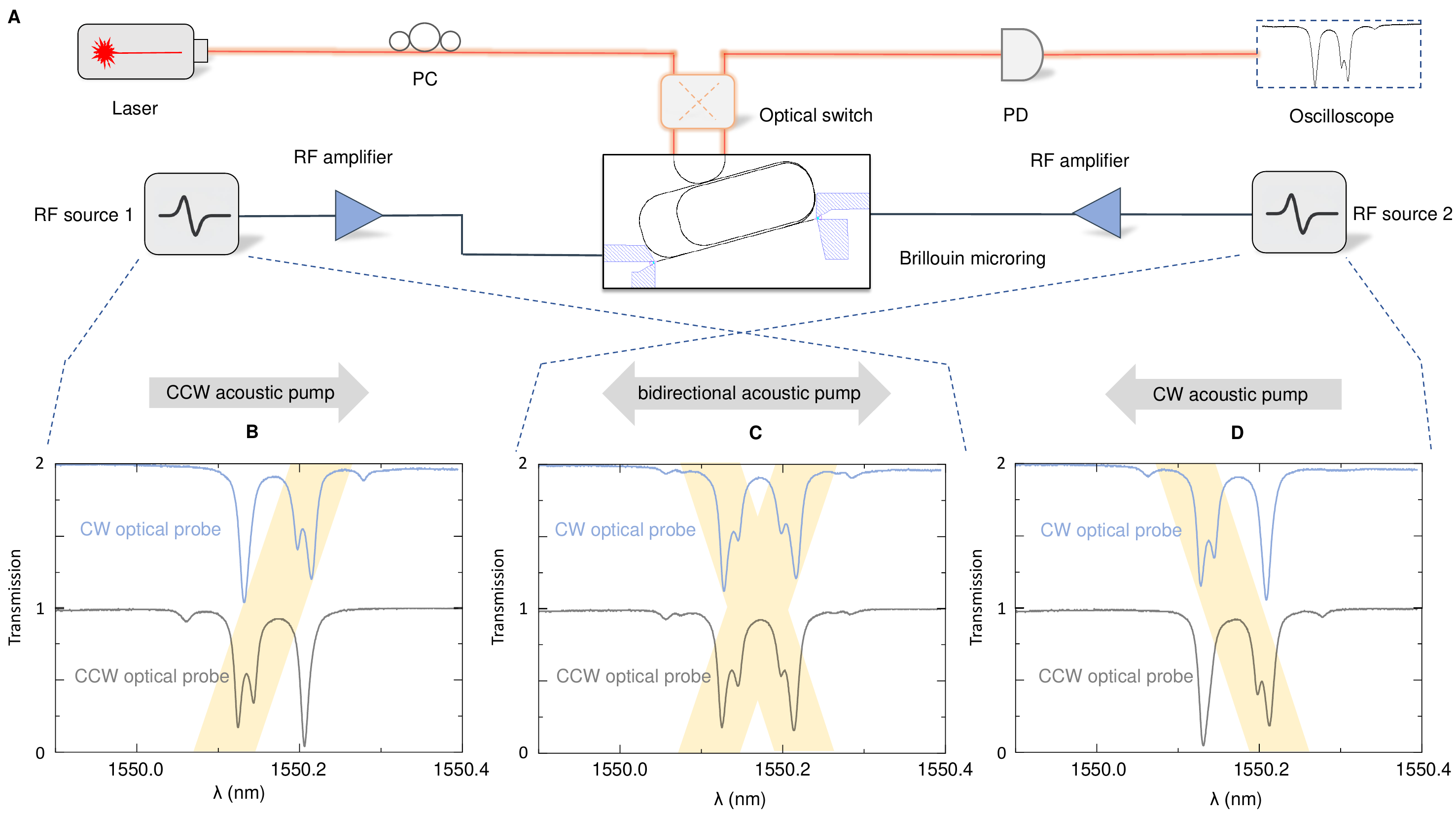}
\caption{\textbf{Multi-degree-of-freedom control of optical microring.} (A) Schematic of the experimental setup. A tunable laser probe is routed via a polarization controller (PC) and an optical switch, which selects the probe direction (CW or CCW). The transmission is measured by a photodetector (PD). Two independent radio-frequency (RF) sources (1 and 2) are amplified and connected to the two on-chip interdigital transducers (IDTs), allowing for independent and directional control of the acoustic pump (CCW, CW, or bidirectional). (B-D) Experimental demonstration of chiral control. The top (blue) and bottom (gray) spectra in each panel represent the transmission for CW and CCW optical probes, respectively. The three columns correspond to different acoustic pump configurations: (B) A single counter-clockwise (CCW) acoustic pump. (D) A single clockwise (CW) acoustic pump. (C) A bidirectional acoustic pump. In all cases, the applied acoustic power is 1 mW. For the bidirectional case (C), the two acoustic pumps are detuned by 1 MHz to avoid the formation of an acoustic Fabry-Perot (F-P) cavity.}
\label{Fig3}
\end{figure*}

\begin{figure*}[t]
\includegraphics[width=1\textwidth]{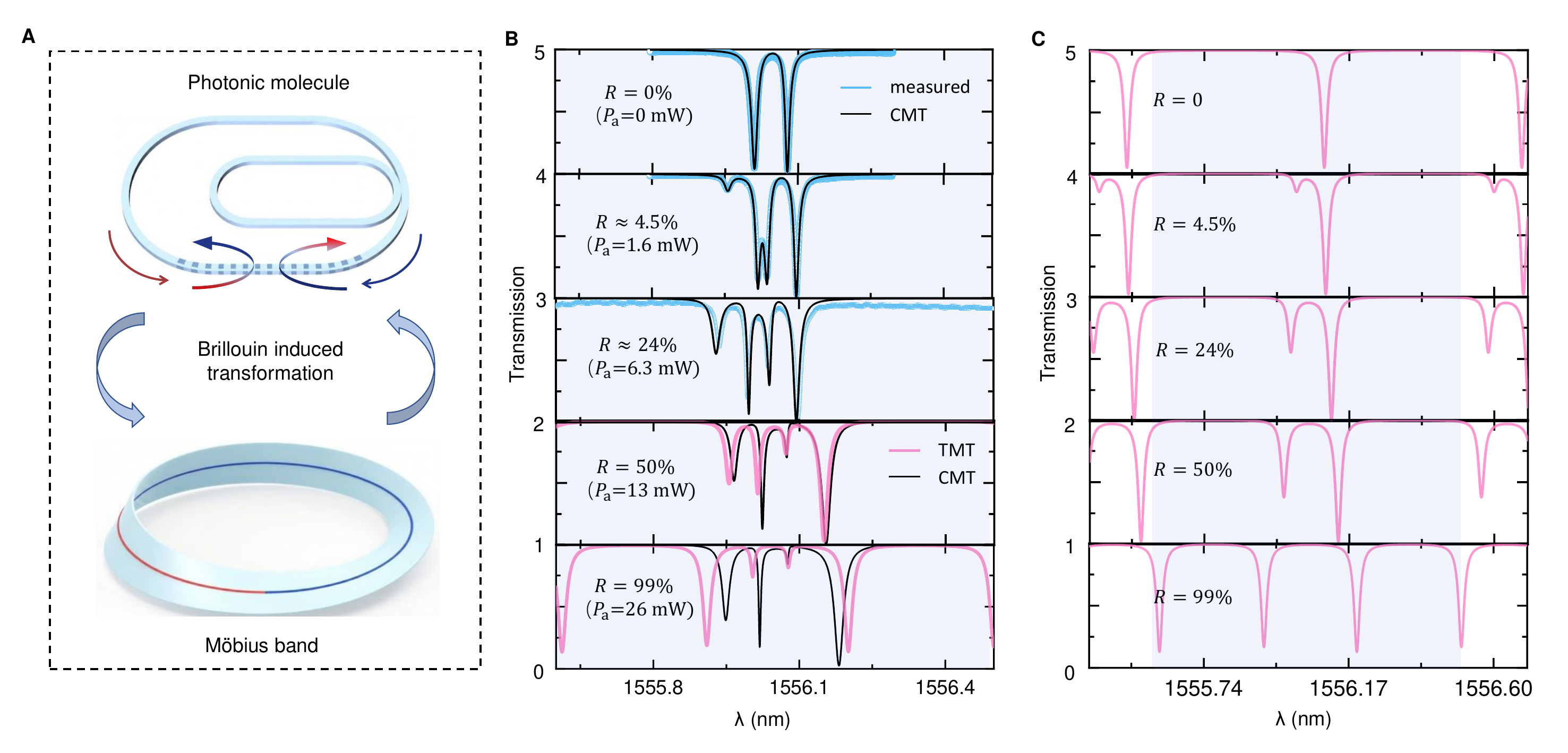}
\caption{\textbf{Light path topology: from a photonic molecule to a M\"obius band.} (A) Schematic of the system's optical path topology. Top: The ``photonic molecule" path, where the outer large ring and the inner small ring are coupled, forming hybrid supermodes. Bottom: The ``M\"obius strip" path. The optical field must complete two round trips to form a cloesd resonant path due to the DBM with high reflectivity. (B) Transmission spectra under strong acoustic pump. The top three traces are experimental measurements (solid blue lines) and predictions using CMT (black lines), showing the complex spectral response. The difference between the two indicates the gradual breakdown of CMT. The bottom two traces are theoretical predictions using CMT (black lines) and TMT (red lines) . (C) Theoretically calculated transmission spectra for a hypothetical system, assuming the outer and inner rings are completely decoupled. The background colors in (B) and (C) cover the same wavelength range.
}
\label{Fig4}
\end{figure*}

The reconfiguration of the microring is systematically characterized by the transmission spectra (CW probe) at varying acoustic powers, as shown in Fig.~\ref{Fig2}(A). At a low power ($P_{\text{a}}=0.2 \ \text{mW}$), the supermode at longer wavelength (Region II) displays obvious broadening, indicating the acoustically-stimulated Brillouin interact between this supermode and the corresponding counter-propagating supermode. As $P_{\text{a}}$ is increased, a splitting of the resonances emerges, manifesting the strong coupling between the optical modes through acoustical drive. Using input-output theory, we predict transmission spectra across all drive powers using a single global fitting parameter $\beta/\sqrt{P_{\mathrm{a}}}=2\pi\times 1.33\,\mathrm{GHz/\sqrt{mW}}$. We found that the theoretical model (dashed lines) accurately captures the observed spectral shapes throughout the measured power range ($P_\mathrm{a}\leq 2.5\,\mathrm{mW}$), validating our understanding of the system dynamics.

To quantify the interaction strength, we introduce the acoustically-stimulated Brillouin interaction cooperativity  $C = |G|^2/\kappa_1\kappa_2$ as a a key figure of merit for coherent interactions, with $\kappa_{2(1)}/2\pi=0.93(0.69)\,\mathrm{GHz}$ are the amplitude decay rates of the two supermodes, with subscript 1(2) denoting the shorter(longer) wavelength supermode in Region I(II). We determine a normalized cooperativity of $C/P_\mathrm{a} = 2.46 \ \text{mW}^{-1}$, corresponding to $C=6.2$ for $P_\mathrm{a}=2.5\,\mathrm{mW}$. This confirms operation deep within the strong coupling regime that coherent energy exchange between optical supermodes before dissipated, leading to the significant splitting spectra in the frequency domain.

The characteristic of the DBM-induced optical mode strong coupling are further quantified in Fig.~\ref{Fig2}(B). The normalized mode splitting $\Delta f_2/\kappa_{2}$ shows square-root dependence on acoustic power that agree with prediction, confirming that the coupling strength $\beta\propto\sqrt{P}$ that scales with acoustic field amplitude. The temporal response of the DBM is investigated in Fig.~\ref{Fig2}(C). By scanning the acoustic drive frequency $\Omega$, we found that the mode splitting is maximized at zero detuning and decreases smoothly as the acoustic frequency is tuned away from resonance. Since the Brillouin interaction phase matching condition, the slight change of $\Omega$ only introduce negligible detuning at optical frequencies, thus there is no avoid-crossing-like spectral feature. Instead, the phase-matching give rise to the shift of working wavelength on $\Omega$ as $\sim 0.17\,\mathrm{nm/MHz}$~\cite{Wang2025}, give rise to a finite DBM bandwidth described by $|\beta|\propto 1/\sqrt{(1+(\delta/(v_g\gamma))^2)}$, where $\delta$, $v_g$, and $\gamma$ represent the detuning, group velocity, and transmission loss of the acoustic wave, respectively. The finite bandwidth leads to the finite temporal response of the system, as shown in the inset: a fixed probe on-resonance with the mode in Region II exhibits a response time of $\sim 300\,\mathrm{ns}$ with respect to a switching of 1\,mW acoustic drive. Figures~\ref{Fig2}(D) and (E) characterize the off-resonant coupling, as the emergence of a resonance dip in Region III and the frequency shift in Region I. Both the extinction ratio (ER) of the new mode and the detuning ($\Delta f_{13}$) between the resonances in Regions I and III showing approximately linear dependence on the parameter $\beta^2/\Omega\propto P_\mathrm{a}$ for $\beta\ll\Omega$ with the perturbative largely detuned approximation, rather than  the square-root dependence.

Beyond the control knobs based on the acoustic drive power and frequency detuning, the DBM provides an additional degree of freedom through pump direction. As shown by the experimental setup in Fig.~\ref{Fig3}(A), the two distinct acoustic ports allow independent control of the CW (via RF Source 2) and CCW (via RF Source 1) acoustic drives, respectively. Figures~\ref{Fig3}(B)-(D) demonstrates three distinct operational regimes programmed by the acoustic drive configuration. In Figs.~\ref{Fig3}(B) and (D), activating only the CCW or CW acoustic pump produces a asymmetric optical response, but showing interesting symmetric between CW and CCW probes: CCW drive and CW probe generate spectrum almost the same as CW drive and CCW probe. The symmetry is also originate from the symmetry in Hamiltonian by exchange all CW and CCW subscripts.
In Fig.~\ref{Fig3}(B), simultaneous activation of both acoustic ports launches two independent counter-propagating traveling waves, enabling concurrent control of both directions. Notably, this bidirectional configuration is not a standing wave but rather two independently controllable interactions, as evidenced by the simultaneous splitting observed in both CW and CCW probe spectra with magnitudes determined by their respective contra-directional acoustic drives. These results demonstrate a fully programmable acousto-optic interface where the two acoustic ports act as independent degrees of freedom for precisely tuning the coupling strengths of two orthogonal mode sets.

When further increase the acoustic drive power, the photonic molecule picture becomes invalid if the acoustically-induced coupling strength $\beta$ becomes comparable with $g$, corresponding to a drive power $P_\mathrm{a}>5.5\,\mathrm{mW}$. Then, the simplified supermodes and on-resonant strong coupling in Eq.~(\ref{Eq2}) is inappropriate, and we should resort the the Hamiltonian Eq.~(\ref{Eq1}) of CMT for accurate prediction. However, when even further increase the coupling strength, the coupling strength can even be comparable with FSR, leading to the breakdown of the perturbative treatment of the CMT. The physics in this regime demands a fundamentally different perspective: when the DBM reflectivity reaches $R = \mathcal{O}(10\%)$, substantial mixing occurs between the CW and CCW optical paths within the outer MRR. This mixing profoundly alters the optical response of the resonance, leading to the breakdown of both photonic molecule picture and perturbative approximation. As $R$ approaches unity, the DBM effectively cuts the ring topology of the optical path, transforming the simple loop into a M\"{o}bius band-like optical path.

The M\"{o}bius topology can be understood through the following light propagation sequence: a CW signal at frequency $\omega$ travels one round trip in the outer MRR before encountering the DBM, which reflects it as a frequency-shifted CCW signal at $\omega + \Omega$ via anti-Stokes scattering. This CCW signal then propagates for another round trip backwardly, and then it interacts with the DBM through Stokes scattering, regenerating a CW signal at the original frequency $\omega$. Thus, the optical field must complete two physical round trips to form a closed resonant path, which manifest the characteristic of M\"{o}bius topology. This doubling of the effective cavity length fundamentally restructures the optical resonance, for instance, leads to a reduced free spectral range (FSR) to the half of original MRR~\cite{Xu2019,Lin2022,Chen2023}. A crucial consequence of this topological reconfiguration is the effective decoupling of the inner and outer MRR. In the M\"{o}bius regime, the strong CW-CCW coupling in the outer MRR dominates the system dynamics, reducing the inter-MRR coupling $g$ to a perturbative correction. As schematically illustrated in Fig.~\ref{Fig4}(A), the photonic molecule transforms into a single M\"{o}bius-band microring resonator with the inner ring providing only minor spectral modifications.

Figures~\ref{Fig4}(B) presents the experimental CW transmission spectra at varying $P_\mathrm{a}$, compared with theoretical predictions. The largest acoustic drive power of $P_\mathrm{a} = 6.3\,\mathrm{mW}$ is achieved, corresponding to a DBM reflectivity $R = 24\%$. At such a large reflectivity, the splitting of the supermode becomes strongly asymmetric, showing the breakdown of the photonic molecule picture, given the corresponding strong coupling with $\beta/\Omega = 0.57$ and cooperativity $C = 15.5$. We also found that the CMT prediction show obvious deviations from experimental results, in particular for the left-most emerging mode, although excellent agreement have been achieved with $P_\mathrm{a}<2.5\,\mathrm{mW}$.

Due to power handling limitations of our aluminum IDTs, direct experimental characterization beyond $P_\mathrm{a} = 6.3\,\mathrm{mW}$ was not possible. We therefore present theoretical predictions for $R = 50\%$ and $R = 99\%$, anticipated at $P_\mathrm{a} \approx 13\,\mathrm{mW}$ and $P_\mathrm{a} \approx 26\,\mathrm{mW}$, respectively. Given the invalidity of CMT in this regime, we employ non-perturbative transfer matrix theory (TMT) to predict the system behavior, as shown by the red shaded lines, treating the DBM as a discrete optical element with reflectivity $R$ embedded within the outer MRR. This discrepancy between the CMT and TMT unambiguously signals the breakdown of perturbative theory. Additionally, we predicted the spectra for an isolated outer MRR by TMT, i.e., set $g=0$ for the calculations in Fig.~\ref{Fig4}(C). By comparing the situations with coupled MRRs and isolated MRRs, we found the low drive power rigme the system behavior is photonic molecule and the DBM is a perturbation. In contrast, the higher driver power with large $R$ reveals the feature of M\"{o}bius topology: a halved FSR arising from the doubled effective optical path length~\cite{Xu2019,Lin2022,Chen2023}. In this high-reflectivity limit, the spectra of the coupled and isolated systems converge, confirming that the inner MRR contributes only a minor perturbation to the dominant M\"{o}bius-band resonator physics. These experimental observations and numerical analyses provide unambiguous evidence for a dynamically controlled topological transition from a conventional photonic molecule to a M\"{o}bius band configuration driven by the acoustically-induced DBM.


\smallskip{}

\noindent \textbf{\large{}Discussion}{\large\par}

\noindent
Our demonstration of acoustically-driven reconfiguration in integrated optical MRRs establishes a new dimension for dynamic control in photonic integrated circuits. Our experimental demonstrations rely on the Zhengfu architecture, which allows the simultaneous confinement of acoustic and optical waveguides and efficient backward Brillouin scattering interaction, while allows direct and efficient access to the acoustic waveguide through piezoelectric actuation. Unlike conventional tuning mechanisms that modify refractive index or absorption, our approach effectively inserts a DBM into the traveling-wave microring resonator and thus fundamentally change the optical path topology. The DBM is electrically switchable, frequency-tunable, and directionally reversible, with the induced optical mode strong coupling achieved an normalized cooperativity $C/P = 2.46\,\mathrm{mW}^{-1}$, providing an extensible and mW-level drive power for efficient reconfiguration of optical MRR.

Remarkably, we observe a cross-over of MRR topology from photonic molecule to M\"{o}bius strip configuration, demonstrating that the new approach to directly change optical pathways other than tuning the optical amplitude and phase. While our current experiments are constrained by IDT power handling to intermediate coupling strengths, it is potential to have stronger reflectivity by cascading multiple DBMs. The programmable DBM opens pathways toward reconfigurable photonic systems of unprecedented flexibility, realizing fully reconfigurable on-chip photonic networks for signal routing, spectral filtering, and arbitrary cavity resonances synthesize. Additionally, the suspension-free platform promises further extension of the hybrid photonic and phononic circuits~\cite{yang_stimulated_2023} for more complicated functionalities, promise a broader deployment of Brillouin-based photonic technologies~\cite{Wang2025} and potential quantum hybridized circuits~\cite{yang2025multi}.


\smallskip{}
\noindent \textbf{\large{}Acknowledgment}{\large\par}
\noindent This work was funded by the Quantum Science and Technology-National Science and Technology Major Project (Grant Nos.~2024ZD0301500 and 2021ZD0300200), the National Natural Science Foundation of China (Grants No.~92265210, 123B2068, 12504454, 92165209, 92365301, 12374361, 92565301, 12104441). This work is also supported by the Fundamental Research Funds for the Central Universities, the USTC Research Funds of the Double First-Class Initiative, the supercomputing system in the Supercomputing Center of USTC the USTC Center for Micro and Nanoscale Research and Fabrication.


\begin{thebibliography}{48}%
\makeatletter
\providecommand \@ifxundefined [1]{%
 \@ifx{#1\undefined}
}%
\providecommand \@ifnum [1]{%
 \ifnum #1\expandafter \@firstoftwo
 \else \expandafter \@secondoftwo
 \fi
}%
\providecommand \@ifx [1]{%
 \ifx #1\expandafter \@firstoftwo
 \else \expandafter \@secondoftwo
 \fi
}%
\providecommand \natexlab [1]{#1}%
\providecommand \enquote  [1]{``#1''}%
\providecommand \bibnamefont  [1]{#1}%
\providecommand \bibfnamefont [1]{#1}%
\providecommand \citenamefont [1]{#1}%
\providecommand \href@noop [0]{\@secondoftwo}%
\providecommand \href [0]{\begingroup \@sanitize@url \@href}%
\providecommand \@href[1]{\@@startlink{#1}\@@href}%
\providecommand \@@href[1]{\endgroup#1\@@endlink}%
\providecommand \@sanitize@url [0]{\catcode `\\12\catcode `\$12\catcode
  `\&12\catcode `\#12\catcode `\^12\catcode `\_12\catcode `\%12\relax}%
\providecommand \@@startlink[1]{}%
\providecommand \@@endlink[0]{}%
\providecommand \url  [0]{\begingroup\@sanitize@url \@url }%
\providecommand \@url [1]{\endgroup\@href {#1}{\urlprefix }}%
\providecommand \urlprefix  [0]{URL }%
\providecommand \Eprint [0]{\href }%
\providecommand \doibase [0]{http://dx.doi.org/}%
\providecommand \selectlanguage [0]{\@gobble}%
\providecommand \bibinfo  [0]{\@secondoftwo}%
\providecommand \bibfield  [0]{\@secondoftwo}%
\providecommand \translation [1]{[#1]}%
\providecommand \BibitemOpen [0]{}%
\providecommand \bibitemStop [0]{}%
\providecommand \bibitemNoStop [0]{.\EOS\space}%
\providecommand \EOS [0]{\spacefactor3000\relax}%
\providecommand \BibitemShut  [1]{\csname bibitem#1\endcsname}%
\let\auto@bib@innerbib\@empty
\bibitem [{\citenamefont {Bogaerts}\ \emph {et~al.}(2020)\citenamefont
  {Bogaerts}, \citenamefont {P{\'e}rez}, \citenamefont {Capmany}, \citenamefont
  {Miller}, \citenamefont {Poon}, \citenamefont {Englund}, \citenamefont
  {Morichetti},\ and\ \citenamefont {Melloni}}]{bogaerts2020programmable}%
  \BibitemOpen
  \bibfield  {author} {\bibinfo {author} {\bibfnamefont {W.}~\bibnamefont
  {Bogaerts}}, \bibinfo {author} {\bibfnamefont {D.}~\bibnamefont {P{\'e}rez}},
  \bibinfo {author} {\bibfnamefont {J.}~\bibnamefont {Capmany}}, \bibinfo
  {author} {\bibfnamefont {D.~A.}\ \bibnamefont {Miller}}, \bibinfo {author}
  {\bibfnamefont {J.}~\bibnamefont {Poon}}, \bibinfo {author} {\bibfnamefont
  {D.}~\bibnamefont {Englund}}, \bibinfo {author} {\bibfnamefont
  {F.}~\bibnamefont {Morichetti}}, \ and\ \bibinfo {author} {\bibfnamefont
  {A.}~\bibnamefont {Melloni}},\ }\bibfield  {title} {\enquote {\bibinfo
  {title} {Programmable photonic circuits},}\ }\href {\doibase
  10.1038/s41586-020-2764-0} {\bibfield  {journal} {\bibinfo  {journal}
  {Nature}\ }\textbf {\bibinfo {volume} {586}},\ \bibinfo {pages} {207}
  (\bibinfo {year} {2020})}\BibitemShut {NoStop}%
\bibitem [{\citenamefont {Marpaung}\ \emph {et~al.}(2019)\citenamefont
  {Marpaung}, \citenamefont {Yao},\ and\ \citenamefont
  {Capmany}}]{marpaung2019integrated}%
  \BibitemOpen
  \bibfield  {author} {\bibinfo {author} {\bibfnamefont {D.}~\bibnamefont
  {Marpaung}}, \bibinfo {author} {\bibfnamefont {J.}~\bibnamefont {Yao}}, \
  and\ \bibinfo {author} {\bibfnamefont {J.}~\bibnamefont {Capmany}},\
  }\bibfield  {title} {\enquote {\bibinfo {title} {Integrated microwave
  photonics},}\ }\href {\doibase 10.1038/s41566-018-0310-5} {\bibfield
  {journal} {\bibinfo  {journal} {Nat. Photon.}\ }\textbf {\bibinfo {volume}
  {13}},\ \bibinfo {pages} {80} (\bibinfo {year} {2019})}\BibitemShut {NoStop}%
\bibitem [{\citenamefont {Boes}\ \emph {et~al.}(2023)\citenamefont {Boes},
  \citenamefont {Chang}, \citenamefont {Langrock}, \citenamefont {Yu},
  \citenamefont {Zhang}, \citenamefont {Lin}, \citenamefont {Lon{\v{c}}ar},
  \citenamefont {Fejer}, \citenamefont {Bowers},\ and\ \citenamefont
  {Mitchell}}]{boes2023lithium}%
  \BibitemOpen
  \bibfield  {author} {\bibinfo {author} {\bibfnamefont {A.}~\bibnamefont
  {Boes}}, \bibinfo {author} {\bibfnamefont {L.}~\bibnamefont {Chang}},
  \bibinfo {author} {\bibfnamefont {C.}~\bibnamefont {Langrock}}, \bibinfo
  {author} {\bibfnamefont {M.}~\bibnamefont {Yu}}, \bibinfo {author}
  {\bibfnamefont {M.}~\bibnamefont {Zhang}}, \bibinfo {author} {\bibfnamefont
  {Q.}~\bibnamefont {Lin}}, \bibinfo {author} {\bibfnamefont {M.}~\bibnamefont
  {Lon{\v{c}}ar}}, \bibinfo {author} {\bibfnamefont {M.}~\bibnamefont {Fejer}},
  \bibinfo {author} {\bibfnamefont {J.}~\bibnamefont {Bowers}}, \ and\ \bibinfo
  {author} {\bibfnamefont {A.}~\bibnamefont {Mitchell}},\ }\bibfield  {title}
  {\enquote {\bibinfo {title} {Lithium niobate photonics: Unlocking the
  electromagnetic spectrum},}\ }\href {\doibase 10.1126/science.abj4396}
  {\bibfield  {journal} {\bibinfo  {journal} {Science}\ }\textbf {\bibinfo
  {volume} {379}},\ \bibinfo {pages} {eabj4396} (\bibinfo {year}
  {2023})}\BibitemShut {NoStop}%
\bibitem [{\citenamefont {Shekhar}\ \emph {et~al.}(2024)\citenamefont
  {Shekhar}, \citenamefont {Bogaerts}, \citenamefont {Chrostowski},
  \citenamefont {Bowers}, \citenamefont {Hochberg}, \citenamefont {Soref},\
  and\ \citenamefont {Shastri}}]{Shekhar2024}%
  \BibitemOpen
  \bibfield  {author} {\bibinfo {author} {\bibfnamefont {S.}~\bibnamefont
  {Shekhar}}, \bibinfo {author} {\bibfnamefont {W.}~\bibnamefont {Bogaerts}},
  \bibinfo {author} {\bibfnamefont {L.}~\bibnamefont {Chrostowski}}, \bibinfo
  {author} {\bibfnamefont {J.~E.}\ \bibnamefont {Bowers}}, \bibinfo {author}
  {\bibfnamefont {M.}~\bibnamefont {Hochberg}}, \bibinfo {author}
  {\bibfnamefont {R.}~\bibnamefont {Soref}}, \ and\ \bibinfo {author}
  {\bibfnamefont {B.~J.}\ \bibnamefont {Shastri}},\ }\bibfield  {title}
  {\enquote {\bibinfo {title} {{Roadmapping the next generation of silicon
  photonics}},}\ }\href {\doibase 10.1038/s41467-024-44750-0} {\bibfield
  {journal} {\bibinfo  {journal} {Nat. Commun.}\ }\textbf {\bibinfo {volume}
  {15}},\ \bibinfo {pages} {751} (\bibinfo {year} {2024})}\BibitemShut
  {NoStop}%
\bibitem [{\citenamefont {Wabnitz}\ \emph {et~al.}(2015)\citenamefont
  {Wabnitz}, \citenamefont {Eggleton} \emph {et~al.}}]{wabnitz2015all}%
  \BibitemOpen
  \bibfield  {author} {\bibinfo {author} {\bibfnamefont {S.}~\bibnamefont
  {Wabnitz}}, \bibinfo {author} {\bibfnamefont {B.~J.}\ \bibnamefont
  {Eggleton}},  \emph {et~al.},\ }\href {\doibase 10.1007/978-3-319-14992-9}
  {\emph {\bibinfo {title} {All-optical signal processing}}},\ Vol.\ \bibinfo
  {volume} {194}\ (\bibinfo  {publisher} {Springer},\ \bibinfo {year}
  {2015})\BibitemShut {NoStop}%
\bibitem [{\citenamefont {O'Brien}\ \emph {et~al.}(2009)\citenamefont
  {O'Brien}, \citenamefont {Furusawa},\ and\ \citenamefont
  {Vu{\v{c}}kovi{\'{c}}}}]{OBrien2009}%
  \BibitemOpen
  \bibfield  {author} {\bibinfo {author} {\bibfnamefont {J.~L.}\ \bibnamefont
  {O'Brien}}, \bibinfo {author} {\bibfnamefont {A.}~\bibnamefont {Furusawa}}, \
  and\ \bibinfo {author} {\bibfnamefont {J.}~\bibnamefont
  {Vu{\v{c}}kovi{\'{c}}}},\ }\bibfield  {title} {\enquote {\bibinfo {title}
  {{Photonic quantum technologies}},}\ }\href {\doibase
  10.1038/nphoton.2009.229} {\bibfield  {journal} {\bibinfo  {journal} {Nat.
  Photon.}\ }\textbf {\bibinfo {volume} {3}},\ \bibinfo {pages} {687} (\bibinfo
  {year} {2009})}\BibitemShut {NoStop}%
\bibitem [{\citenamefont {Elshaari}\ \emph {et~al.}(2020)\citenamefont
  {Elshaari}, \citenamefont {Pernice}, \citenamefont {Srinivasan},
  \citenamefont {Benson},\ and\ \citenamefont {Zwiller}}]{Elshaari2020}%
  \BibitemOpen
  \bibfield  {author} {\bibinfo {author} {\bibfnamefont {A.~W.}\ \bibnamefont
  {Elshaari}}, \bibinfo {author} {\bibfnamefont {W.}~\bibnamefont {Pernice}},
  \bibinfo {author} {\bibfnamefont {K.}~\bibnamefont {Srinivasan}}, \bibinfo
  {author} {\bibfnamefont {O.}~\bibnamefont {Benson}}, \ and\ \bibinfo {author}
  {\bibfnamefont {V.}~\bibnamefont {Zwiller}},\ }\bibfield  {title} {\enquote
  {\bibinfo {title} {{Hybrid integrated quantum photonic circuits}},}\ }\href
  {\doibase 10.1038/s41566-020-0609-x} {\bibfield  {journal} {\bibinfo
  {journal} {Nat. Photon.}\ }\textbf {\bibinfo {volume} {14}},\ \bibinfo
  {pages} {285} (\bibinfo {year} {2020})}\BibitemShut {NoStop}%
\bibitem [{\citenamefont {Pelucchi}\ \emph {et~al.}(2022)\citenamefont
  {Pelucchi}, \citenamefont {Fagas}, \citenamefont {Aharonovich}, \citenamefont
  {Englund}, \citenamefont {Figueroa}, \citenamefont {Gong}, \citenamefont
  {Hannes}, \citenamefont {Liu}, \citenamefont {Lu}, \citenamefont {Matsuda},
  \citenamefont {Pan}, \citenamefont {Schreck}, \citenamefont {Sciarrino},
  \citenamefont {Silberhorn}, \citenamefont {Wang},\ and\ \citenamefont
  {Jons}}]{Pelucchi2022}%
  \BibitemOpen
  \bibfield  {author} {\bibinfo {author} {\bibfnamefont {E.}~\bibnamefont
  {Pelucchi}}, \bibinfo {author} {\bibfnamefont {G.}~\bibnamefont {Fagas}},
  \bibinfo {author} {\bibfnamefont {I.}~\bibnamefont {Aharonovich}}, \bibinfo
  {author} {\bibfnamefont {D.}~\bibnamefont {Englund}}, \bibinfo {author}
  {\bibfnamefont {E.}~\bibnamefont {Figueroa}}, \bibinfo {author}
  {\bibfnamefont {Q.}~\bibnamefont {Gong}}, \bibinfo {author} {\bibfnamefont
  {H.}~\bibnamefont {Hannes}}, \bibinfo {author} {\bibfnamefont
  {J.}~\bibnamefont {Liu}}, \bibinfo {author} {\bibfnamefont {C.-Y.}\
  \bibnamefont {Lu}}, \bibinfo {author} {\bibfnamefont {N.}~\bibnamefont
  {Matsuda}}, \bibinfo {author} {\bibfnamefont {J.-W.}\ \bibnamefont {Pan}},
  \bibinfo {author} {\bibfnamefont {F.}~\bibnamefont {Schreck}}, \bibinfo
  {author} {\bibfnamefont {F.}~\bibnamefont {Sciarrino}}, \bibinfo {author}
  {\bibfnamefont {C.}~\bibnamefont {Silberhorn}}, \bibinfo {author}
  {\bibfnamefont {J.}~\bibnamefont {Wang}}, \ and\ \bibinfo {author}
  {\bibfnamefont {K.~D.}\ \bibnamefont {Jons}},\ }\bibfield  {title}
  {\enquote {\bibinfo {title} {{The potential and global outlook of integrated
  photonics for quantum technologies}},}\ }\href {\doibase
  10.1038/s42254-021-00398-z} {\bibfield  {journal} {\bibinfo  {journal} {Nat.
  Rev. Phys.}\ }\textbf {\bibinfo {volume} {4}},\ \bibinfo {pages} {194}
  (\bibinfo {year} {2022})}\BibitemShut {NoStop}%
\bibitem [{\citenamefont {Cheng}\ \emph {et~al.}(2023)\citenamefont {Cheng},
  \citenamefont {Zhou}, \citenamefont {Wang}, \citenamefont {Shen},
  \citenamefont {Taher},\ and\ \citenamefont {Tang}}]{Cheng2023}%
  \BibitemOpen
  \bibfield  {author} {\bibinfo {author} {\bibfnamefont {R.}~\bibnamefont
  {Cheng}}, \bibinfo {author} {\bibfnamefont {Y.}~\bibnamefont {Zhou}},
  \bibinfo {author} {\bibfnamefont {S.}~\bibnamefont {Wang}}, \bibinfo {author}
  {\bibfnamefont {M.}~\bibnamefont {Shen}}, \bibinfo {author} {\bibfnamefont
  {T.}~\bibnamefont {Taher}}, \ and\ \bibinfo {author} {\bibfnamefont {H.~X.}\
  \bibnamefont {Tang}},\ }\bibfield  {title} {\enquote {\bibinfo {title} {{A
  100-pixel photon-number-resolving detector unveiling photon statistics}},}\
  }\href {\doibase 10.1038/s41566-022-01119-3} {\bibfield  {journal} {\bibinfo
  {journal} {Nat. Photon.}\ }\textbf {\bibinfo {volume} {17}},\ \bibinfo
  {pages} {112} (\bibinfo {year} {2023})}\BibitemShut {NoStop}%
\bibitem [{\citenamefont {Liu}\ \emph {et~al.}(2022)\citenamefont {Liu},
  \citenamefont {Bo}, \citenamefont {Chang}, \citenamefont {Dong},
  \citenamefont {Ou}, \citenamefont {Regan}, \citenamefont {Shen},
  \citenamefont {Song}, \citenamefont {Yao}, \citenamefont {Zhang},
  \citenamefont {Zou},\ and\ \citenamefont {Xiao}}]{Liu2022}%
  \BibitemOpen
  \bibfield  {author} {\bibinfo {author} {\bibfnamefont {J.}~\bibnamefont
  {Liu}}, \bibinfo {author} {\bibfnamefont {F.}~\bibnamefont {Bo}}, \bibinfo
  {author} {\bibfnamefont {L.}~\bibnamefont {Chang}}, \bibinfo {author}
  {\bibfnamefont {C.-H.}\ \bibnamefont {Dong}}, \bibinfo {author}
  {\bibfnamefont {X.}~\bibnamefont {Ou}}, \bibinfo {author} {\bibfnamefont
  {B.}~\bibnamefont {Regan}}, \bibinfo {author} {\bibfnamefont
  {X.}~\bibnamefont {Shen}}, \bibinfo {author} {\bibfnamefont {Q.}~\bibnamefont
  {Song}}, \bibinfo {author} {\bibfnamefont {B.}~\bibnamefont {Yao}}, \bibinfo
  {author} {\bibfnamefont {W.}~\bibnamefont {Zhang}}, \bibinfo {author}
  {\bibfnamefont {C.-L.}\ \bibnamefont {Zou}}, \ and\ \bibinfo {author}
  {\bibfnamefont {Y.-F.}\ \bibnamefont {Xiao}},\ }\bibfield  {title} {\enquote
  {\bibinfo {title} {{Emerging material platforms for integrated microcavity
  photonics}},}\ }\href {\doibase 10.1007/s11433-022-1957-3} {\bibfield
  {journal} {\bibinfo  {journal} {Sci. China Phys. Mech.}\ }\textbf {\bibinfo
  {volume} {65}},\ \bibinfo {pages} {104201} (\bibinfo {year}
  {2022})}\BibitemShut {NoStop}%
\bibitem [{\citenamefont {Bogaerts}\ \emph {et~al.}(2012)\citenamefont
  {Bogaerts}, \citenamefont {De~Heyn}, \citenamefont {Van~Vaerenbergh},
  \citenamefont {De~Vos}, \citenamefont {Kumar~Selvaraja}, \citenamefont
  {Claes}, \citenamefont {Dumon}, \citenamefont {Bienstman}, \citenamefont
  {Van~Thourhout},\ and\ \citenamefont {Baets}}]{bogaerts2012silicon}%
  \BibitemOpen
  \bibfield  {author} {\bibinfo {author} {\bibfnamefont {W.}~\bibnamefont
  {Bogaerts}}, \bibinfo {author} {\bibfnamefont {P.}~\bibnamefont {De~Heyn}},
  \bibinfo {author} {\bibfnamefont {T.}~\bibnamefont {Van~Vaerenbergh}},
  \bibinfo {author} {\bibfnamefont {K.}~\bibnamefont {De~Vos}}, \bibinfo
  {author} {\bibfnamefont {S.}~\bibnamefont {Kumar~Selvaraja}}, \bibinfo
  {author} {\bibfnamefont {T.}~\bibnamefont {Claes}}, \bibinfo {author}
  {\bibfnamefont {P.}~\bibnamefont {Dumon}}, \bibinfo {author} {\bibfnamefont
  {P.}~\bibnamefont {Bienstman}}, \bibinfo {author} {\bibfnamefont
  {D.}~\bibnamefont {Van~Thourhout}}, \ and\ \bibinfo {author} {\bibfnamefont
  {R.}~\bibnamefont {Baets}},\ }\bibfield  {title} {\enquote {\bibinfo {title}
  {Silicon microring resonators},}\ }\href {\doibase 10.1002/lpor.201100017}
  {\bibfield  {journal} {\bibinfo  {journal} {Laser Photonics Rev.}\ }\textbf
  {\bibinfo {volume} {6}},\ \bibinfo {pages} {47} (\bibinfo {year}
  {2012})}\BibitemShut {NoStop}%
\bibitem [{\citenamefont {Zhu}\ \emph {et~al.}(2021)\citenamefont {Zhu},
  \citenamefont {Shao}, \citenamefont {Yu}, \citenamefont {Cheng},
  \citenamefont {Desiatov}, \citenamefont {Xin}, \citenamefont {Hu},
  \citenamefont {Holzgrafe}, \citenamefont {Ghosh}, \citenamefont
  {Shams-Ansari} \emph {et~al.}}]{zhu2021integrated}%
  \BibitemOpen
  \bibfield  {author} {\bibinfo {author} {\bibfnamefont {D.}~\bibnamefont
  {Zhu}}, \bibinfo {author} {\bibfnamefont {L.}~\bibnamefont {Shao}}, \bibinfo
  {author} {\bibfnamefont {M.}~\bibnamefont {Yu}}, \bibinfo {author}
  {\bibfnamefont {R.}~\bibnamefont {Cheng}}, \bibinfo {author} {\bibfnamefont
  {B.}~\bibnamefont {Desiatov}}, \bibinfo {author} {\bibfnamefont
  {C.}~\bibnamefont {Xin}}, \bibinfo {author} {\bibfnamefont {Y.}~\bibnamefont
  {Hu}}, \bibinfo {author} {\bibfnamefont {J.}~\bibnamefont {Holzgrafe}},
  \bibinfo {author} {\bibfnamefont {S.}~\bibnamefont {Ghosh}}, \bibinfo
  {author} {\bibfnamefont {A.}~\bibnamefont {Shams-Ansari}},  \emph {et~al.},\
  }\bibfield  {title} {\enquote {\bibinfo {title} {Integrated photonics on
  thin-film lithium niobate},}\ }\href {\doibase 10.1364/AOP.411024} {\bibfield
   {journal} {\bibinfo  {journal} {Adv. Opt. Photon}\ }\textbf {\bibinfo
  {volume} {13}},\ \bibinfo {pages} {242} (\bibinfo {year} {2021})}\BibitemShut
  {NoStop}%
\bibitem [{\citenamefont {Jin}\ \emph {et~al.}(2021)\citenamefont {Jin},
  \citenamefont {Yang}, \citenamefont {Chang}, \citenamefont {Shen},
  \citenamefont {Wang}, \citenamefont {Leal}, \citenamefont {Wu}, \citenamefont
  {Gao}, \citenamefont {Feshali}, \citenamefont {Paniccia} \emph
  {et~al.}}]{jin2021hertz}%
  \BibitemOpen
  \bibfield  {author} {\bibinfo {author} {\bibfnamefont {W.}~\bibnamefont
  {Jin}}, \bibinfo {author} {\bibfnamefont {Q.-F.}\ \bibnamefont {Yang}},
  \bibinfo {author} {\bibfnamefont {L.}~\bibnamefont {Chang}}, \bibinfo
  {author} {\bibfnamefont {B.}~\bibnamefont {Shen}}, \bibinfo {author}
  {\bibfnamefont {H.}~\bibnamefont {Wang}}, \bibinfo {author} {\bibfnamefont
  {M.~A.}\ \bibnamefont {Leal}}, \bibinfo {author} {\bibfnamefont
  {L.}~\bibnamefont {Wu}}, \bibinfo {author} {\bibfnamefont {M.}~\bibnamefont
  {Gao}}, \bibinfo {author} {\bibfnamefont {A.}~\bibnamefont {Feshali}},
  \bibinfo {author} {\bibfnamefont {M.}~\bibnamefont {Paniccia}},  \emph
  {et~al.},\ }\bibfield  {title} {\enquote {\bibinfo {title} {Hertz-linewidth
  semiconductor lasers using cmos-ready ultra-high-q microresonators},}\ }\href
  {\doibase 10.1038/s41566-021-00761-7} {\bibfield  {journal} {\bibinfo
  {journal} {Nat. Photon.}\ }\textbf {\bibinfo {volume} {15}},\ \bibinfo
  {pages} {346} (\bibinfo {year} {2021})}\BibitemShut {NoStop}%
\bibitem [{\citenamefont {Liu}\ \emph {et~al.}(2024)\citenamefont {Liu},
  \citenamefont {Qiu}, \citenamefont {Ji}, \citenamefont {Bancora},
  \citenamefont {Lihachev}, \citenamefont {Riemensberger}, \citenamefont
  {Wang}, \citenamefont {Voloshin},\ and\ \citenamefont
  {Kippenberg}}]{liu2024fully}%
  \BibitemOpen
  \bibfield  {author} {\bibinfo {author} {\bibfnamefont {Y.}~\bibnamefont
  {Liu}}, \bibinfo {author} {\bibfnamefont {Z.}~\bibnamefont {Qiu}}, \bibinfo
  {author} {\bibfnamefont {X.}~\bibnamefont {Ji}}, \bibinfo {author}
  {\bibfnamefont {A.}~\bibnamefont {Bancora}}, \bibinfo {author} {\bibfnamefont
  {G.}~\bibnamefont {Lihachev}}, \bibinfo {author} {\bibfnamefont
  {J.}~\bibnamefont {Riemensberger}}, \bibinfo {author} {\bibfnamefont {R.~N.}\
  \bibnamefont {Wang}}, \bibinfo {author} {\bibfnamefont {A.}~\bibnamefont
  {Voloshin}}, \ and\ \bibinfo {author} {\bibfnamefont {T.~J.}\ \bibnamefont
  {Kippenberg}},\ }\bibfield  {title} {\enquote {\bibinfo {title} {A fully
  hybrid integrated erbium-based laser},}\ }\href {\doibase
  10.1038/s41566-024-01454-7} {\bibfield  {journal} {\bibinfo  {journal} {Nat.
  Photon.}\ }\textbf {\bibinfo {volume} {18}},\ \bibinfo {pages} {829}
  (\bibinfo {year} {2024})}\BibitemShut {NoStop}%
\bibitem [{\citenamefont {Liu}\ \emph {et~al.}(2021)\citenamefont {Liu},
  \citenamefont {Xu}, \citenamefont {Tan}, \citenamefont {Shi},\ and\
  \citenamefont {Dai}}]{liu2021silicon}%
  \BibitemOpen
  \bibfield  {author} {\bibinfo {author} {\bibfnamefont {D.}~\bibnamefont
  {Liu}}, \bibinfo {author} {\bibfnamefont {H.}~\bibnamefont {Xu}}, \bibinfo
  {author} {\bibfnamefont {Y.}~\bibnamefont {Tan}}, \bibinfo {author}
  {\bibfnamefont {Y.}~\bibnamefont {Shi}}, \ and\ \bibinfo {author}
  {\bibfnamefont {D.}~\bibnamefont {Dai}},\ }\bibfield  {title} {\enquote
  {\bibinfo {title} {Silicon photonic filters},}\ }\href {\doibase
  10.1002/mop.32509} {\bibfield  {journal} {\bibinfo  {journal} {Microw. Opt.
  Technol. Lett.}\ }\textbf {\bibinfo {volume} {63}},\ \bibinfo {pages} {2252}
  (\bibinfo {year} {2021})}\BibitemShut {NoStop}%
\bibitem [{\citenamefont {Xu}\ \emph {et~al.}(2005)\citenamefont {Xu},
  \citenamefont {Schmidt}, \citenamefont {Pradhan},\ and\ \citenamefont
  {Lipson}}]{xu2005micrometre}%
  \BibitemOpen
  \bibfield  {author} {\bibinfo {author} {\bibfnamefont {Q.}~\bibnamefont
  {Xu}}, \bibinfo {author} {\bibfnamefont {B.}~\bibnamefont {Schmidt}},
  \bibinfo {author} {\bibfnamefont {S.}~\bibnamefont {Pradhan}}, \ and\
  \bibinfo {author} {\bibfnamefont {M.}~\bibnamefont {Lipson}},\ }\bibfield
  {title} {\enquote {\bibinfo {title} {Micrometre-scale silicon electro-optic
  modulator},}\ }\href {\doibase 10.1038/nature03569} {\bibfield  {journal}
  {\bibinfo  {journal} {Nature}\ }\textbf {\bibinfo {volume} {435}},\ \bibinfo
  {pages} {325} (\bibinfo {year} {2005})}\BibitemShut {NoStop}%
\bibitem [{\citenamefont {Wang}\ \emph {et~al.}(2018)\citenamefont {Wang},
  \citenamefont {Zhang}, \citenamefont {Chen}, \citenamefont {Bertrand},
  \citenamefont {Shams-Ansari}, \citenamefont {Chandrasekhar}, \citenamefont
  {Winzer},\ and\ \citenamefont {Lon{\v{c}}ar}}]{wang2018integrated}%
  \BibitemOpen
  \bibfield  {author} {\bibinfo {author} {\bibfnamefont {C.}~\bibnamefont
  {Wang}}, \bibinfo {author} {\bibfnamefont {M.}~\bibnamefont {Zhang}},
  \bibinfo {author} {\bibfnamefont {X.}~\bibnamefont {Chen}}, \bibinfo {author}
  {\bibfnamefont {M.}~\bibnamefont {Bertrand}}, \bibinfo {author}
  {\bibfnamefont {A.}~\bibnamefont {Shams-Ansari}}, \bibinfo {author}
  {\bibfnamefont {S.}~\bibnamefont {Chandrasekhar}}, \bibinfo {author}
  {\bibfnamefont {P.}~\bibnamefont {Winzer}}, \ and\ \bibinfo {author}
  {\bibfnamefont {M.}~\bibnamefont {Lon{\v{c}}ar}},\ }\bibfield  {title}
  {\enquote {\bibinfo {title} {Integrated lithium niobate electro-optic
  modulators operating at cmos-compatible voltages},}\ }\href {\doibase
  10.1038/s41586-018-0551-y} {\bibfield  {journal} {\bibinfo  {journal}
  {Nature}\ }\textbf {\bibinfo {volume} {562}},\ \bibinfo {pages} {101}
  (\bibinfo {year} {2018})}\BibitemShut {NoStop}%
\bibitem [{\citenamefont {Zhang}\ \emph
  {et~al.}(2019{\natexlab{a}})\citenamefont {Zhang}, \citenamefont {Buscaino},
  \citenamefont {Wang}, \citenamefont {Shams-Ansari}, \citenamefont {Reimer},
  \citenamefont {Zhu}, \citenamefont {Kahn},\ and\ \citenamefont
  {Lon{\v{c}}ar}}]{zhang2019broadband}%
  \BibitemOpen
  \bibfield  {author} {\bibinfo {author} {\bibfnamefont {M.}~\bibnamefont
  {Zhang}}, \bibinfo {author} {\bibfnamefont {B.}~\bibnamefont {Buscaino}},
  \bibinfo {author} {\bibfnamefont {C.}~\bibnamefont {Wang}}, \bibinfo {author}
  {\bibfnamefont {A.}~\bibnamefont {Shams-Ansari}}, \bibinfo {author}
  {\bibfnamefont {C.}~\bibnamefont {Reimer}}, \bibinfo {author} {\bibfnamefont
  {R.}~\bibnamefont {Zhu}}, \bibinfo {author} {\bibfnamefont {J.~M.}\
  \bibnamefont {Kahn}}, \ and\ \bibinfo {author} {\bibfnamefont
  {M.}~\bibnamefont {Lon{\v{c}}ar}},\ }\bibfield  {title} {\enquote {\bibinfo
  {title} {Broadband electro-optic frequency comb generation in a lithium
  niobate microring resonator},}\ }\href {\doibase 10.1038/s41586-019-1008-7}
  {\bibfield  {journal} {\bibinfo  {journal} {Nature}\ }\textbf {\bibinfo
  {volume} {568}},\ \bibinfo {pages} {373} (\bibinfo {year}
  {2019}{\natexlab{a}})}\BibitemShut {NoStop}%
\bibitem [{\citenamefont {Gaeta}\ \emph {et~al.}(2019)\citenamefont {Gaeta},
  \citenamefont {Lipson},\ and\ \citenamefont {Kippenberg}}]{Gaeta2019}%
  \BibitemOpen
  \bibfield  {author} {\bibinfo {author} {\bibfnamefont {A.~L.}\ \bibnamefont
  {Gaeta}}, \bibinfo {author} {\bibfnamefont {M.}~\bibnamefont {Lipson}}, \
  and\ \bibinfo {author} {\bibfnamefont {T.~J.}\ \bibnamefont {Kippenberg}},\
  }\bibfield  {title} {\enquote {\bibinfo {title} {{Photonic-chip-based
  frequency combs}},}\ }\href {\doibase 10.1038/s41566-019-0358-x} {\bibfield
  {journal} {\bibinfo  {journal} {Nat. Photon.}\ }\textbf {\bibinfo {volume}
  {13}},\ \bibinfo {pages} {158} (\bibinfo {year} {2019})}\BibitemShut
  {NoStop}%
\bibitem [{\citenamefont {Chang}\ \emph {et~al.}(2022)\citenamefont {Chang},
  \citenamefont {Liu},\ and\ \citenamefont {Bowers}}]{Chang2022}%
  \BibitemOpen
  \bibfield  {author} {\bibinfo {author} {\bibfnamefont {L.}~\bibnamefont
  {Chang}}, \bibinfo {author} {\bibfnamefont {S.}~\bibnamefont {Liu}}, \ and\
  \bibinfo {author} {\bibfnamefont {J.~E.}\ \bibnamefont {Bowers}},\ }\bibfield
   {title} {\enquote {\bibinfo {title} {{Integrated optical frequency comb
  technologies}},}\ }\href {\doibase 10.1038/s41566-021-00945-1} {\bibfield
  {journal} {\bibinfo  {journal} {Nat. Photon.}\ }\textbf {\bibinfo {volume}
  {16}},\ \bibinfo {pages} {95} (\bibinfo {year} {2022})}\BibitemShut {NoStop}%
\bibitem [{\citenamefont {Ramachandran}\ \emph {et~al.}(2008)\citenamefont
  {Ramachandran}, \citenamefont {Wang}, \citenamefont {Clarke}, \citenamefont
  {Ja}, \citenamefont {Goad}, \citenamefont {Wald}, \citenamefont {Flood},
  \citenamefont {Knobbe}, \citenamefont {Hryniewicz}, \citenamefont {Chu} \emph
  {et~al.}}]{ramachandran2008universal}%
  \BibitemOpen
  \bibfield  {author} {\bibinfo {author} {\bibfnamefont {A.}~\bibnamefont
  {Ramachandran}}, \bibinfo {author} {\bibfnamefont {S.}~\bibnamefont {Wang}},
  \bibinfo {author} {\bibfnamefont {J.}~\bibnamefont {Clarke}}, \bibinfo
  {author} {\bibfnamefont {S.}~\bibnamefont {Ja}}, \bibinfo {author}
  {\bibfnamefont {D.}~\bibnamefont {Goad}}, \bibinfo {author} {\bibfnamefont
  {L.}~\bibnamefont {Wald}}, \bibinfo {author} {\bibfnamefont {E.}~\bibnamefont
  {Flood}}, \bibinfo {author} {\bibfnamefont {E.}~\bibnamefont {Knobbe}},
  \bibinfo {author} {\bibfnamefont {J.}~\bibnamefont {Hryniewicz}}, \bibinfo
  {author} {\bibfnamefont {S.}~\bibnamefont {Chu}},  \emph {et~al.},\
  }\bibfield  {title} {\enquote {\bibinfo {title} {A universal biosensing
  platform based on optical micro-ring resonators},}\ }\href {\doibase
  10.1016/j.bios.2007.09.007} {\bibfield  {journal} {\bibinfo  {journal}
  {Biosens. Bioelectron.}\ }\textbf {\bibinfo {volume} {23}},\ \bibinfo {pages}
  {939} (\bibinfo {year} {2008})}\BibitemShut {NoStop}%
\bibitem [{\citenamefont {Strekalov}\ \emph {et~al.}(2016)\citenamefont
  {Strekalov}, \citenamefont {Marquardt}, \citenamefont {Matsko}, \citenamefont
  {Schwefel},\ and\ \citenamefont {Leuchs}}]{strekalov2016nonlinear}%
  \BibitemOpen
  \bibfield  {author} {\bibinfo {author} {\bibfnamefont {D.~V.}\ \bibnamefont
  {Strekalov}}, \bibinfo {author} {\bibfnamefont {C.}~\bibnamefont
  {Marquardt}}, \bibinfo {author} {\bibfnamefont {A.~B.}\ \bibnamefont
  {Matsko}}, \bibinfo {author} {\bibfnamefont {H.~G.}\ \bibnamefont
  {Schwefel}}, \ and\ \bibinfo {author} {\bibfnamefont {G.}~\bibnamefont
  {Leuchs}},\ }\bibfield  {title} {\enquote {\bibinfo {title} {Nonlinear and
  quantum optics with whispering gallery resonators},}\ }\href {\doibase
  10.1088/2040-8978/18/12/123002} {\bibfield  {journal} {\bibinfo  {journal}
  {Journal of Opt.}\ }\textbf {\bibinfo {volume} {18}},\ \bibinfo {pages}
  {123002} (\bibinfo {year} {2016})}\BibitemShut {NoStop}%
\bibitem [{\citenamefont {Eggleton}\ \emph {et~al.}(2019)\citenamefont
  {Eggleton}, \citenamefont {Poulton}, \citenamefont {Rakich}, \citenamefont
  {Steel},\ and\ \citenamefont {Bahl}}]{eggleton2019brillouin}%
  \BibitemOpen
  \bibfield  {author} {\bibinfo {author} {\bibfnamefont {B.~J.}\ \bibnamefont
  {Eggleton}}, \bibinfo {author} {\bibfnamefont {C.~G.}\ \bibnamefont
  {Poulton}}, \bibinfo {author} {\bibfnamefont {P.~T.}\ \bibnamefont {Rakich}},
  \bibinfo {author} {\bibfnamefont {M.~J.}\ \bibnamefont {Steel}}, \ and\
  \bibinfo {author} {\bibfnamefont {G.}~\bibnamefont {Bahl}},\ }\bibfield
  {title} {\enquote {\bibinfo {title} {Brillouin integrated photonics},}\
  }\href {\doibase 10.1038/s41566-019-0498-z} {\bibfield  {journal} {\bibinfo
  {journal} {Nat. Photon.}\ }\textbf {\bibinfo {volume} {13}},\ \bibinfo
  {pages} {664} (\bibinfo {year} {2019})}\BibitemShut {NoStop}%
\bibitem [{\citenamefont {Chang}\ \emph {et~al.}(2018)\citenamefont {Chang},
  \citenamefont {Douglas}, \citenamefont {Gonz{\'{a}}lez-Tudela}, \citenamefont
  {Hung},\ and\ \citenamefont {Kimble}}]{Chang2018}%
  \BibitemOpen
  \bibfield  {author} {\bibinfo {author} {\bibfnamefont {D.~E.}\ \bibnamefont
  {Chang}}, \bibinfo {author} {\bibfnamefont {J.~S.}\ \bibnamefont {Douglas}},
  \bibinfo {author} {\bibfnamefont {A.}~\bibnamefont {Gonz{\'{a}}lez-Tudela}},
  \bibinfo {author} {\bibfnamefont {C.-L.}\ \bibnamefont {Hung}}, \ and\
  \bibinfo {author} {\bibfnamefont {H.~J.}\ \bibnamefont {Kimble}},\ }\bibfield
   {title} {\enquote {\bibinfo {title} {{Colloquium : Quantum matter built from
  nanoscopic lattices of atoms and photons}},}\ }\href {\doibase
  10.1103/RevModPhys.90.031002} {\bibfield  {journal} {\bibinfo  {journal}
  {Rev. Mod. Phys.}\ }\textbf {\bibinfo {volume} {90}},\ \bibinfo {pages}
  {031002} (\bibinfo {year} {2018})}\BibitemShut {NoStop}%
\bibitem [{\citenamefont {Li}\ \emph {et~al.}(2020)\citenamefont {Li},
  \citenamefont {Zhang}, \citenamefont {Tang}, \citenamefont {Dong},
  \citenamefont {Guo},\ and\ \citenamefont {Zou}}]{Li2020}%
  \BibitemOpen
  \bibfield  {author} {\bibinfo {author} {\bibfnamefont {M.}~\bibnamefont
  {Li}}, \bibinfo {author} {\bibfnamefont {Y.-L.}\ \bibnamefont {Zhang}},
  \bibinfo {author} {\bibfnamefont {H.~X.}\ \bibnamefont {Tang}}, \bibinfo
  {author} {\bibfnamefont {C.-H.}\ \bibnamefont {Dong}}, \bibinfo {author}
  {\bibfnamefont {G.-C.}\ \bibnamefont {Guo}}, \ and\ \bibinfo {author}
  {\bibfnamefont {C.-L.}\ \bibnamefont {Zou}},\ }\bibfield  {title} {\enquote
  {\bibinfo {title} {{Photon-Photon Quantum Phase Gate in a Photonic Molecule
  with chi(2) Nonlinearity}},}\ }\href {\doibase
  10.1103/PhysRevApplied.13.044013} {\bibfield  {journal} {\bibinfo  {journal}
  {Phys. Rev. Appl.}\ }\textbf {\bibinfo {volume} {13}},\ \bibinfo {pages}
  {044013} (\bibinfo {year} {2020})}\BibitemShut {NoStop}%
\bibitem [{\citenamefont {Zhou}\ \emph {et~al.}(2023)\citenamefont {Zhou},
  \citenamefont {Tamura}, \citenamefont {Chang},\ and\ \citenamefont
  {Hung}}]{Zhou2023}%
  \BibitemOpen
  \bibfield  {author} {\bibinfo {author} {\bibfnamefont {X.}~\bibnamefont
  {Zhou}}, \bibinfo {author} {\bibfnamefont {H.}~\bibnamefont {Tamura}},
  \bibinfo {author} {\bibfnamefont {T.-H.}\ \bibnamefont {Chang}}, \ and\
  \bibinfo {author} {\bibfnamefont {C.-L.}\ \bibnamefont {Hung}},\ }\bibfield
  {title} {\enquote {\bibinfo {title} {{Coupling Single Atoms to a Nanophotonic
  Whispering-Gallery-Mode Resonator via Optical Guiding}},}\ }\href {\doibase
  10.1103/PhysRevLett.130.103601} {\bibfield  {journal} {\bibinfo  {journal}
  {Phys. Rev. Lett.}\ }\textbf {\bibinfo {volume} {130}},\ \bibinfo {pages}
  {103601} (\bibinfo {year} {2023})}\BibitemShut {NoStop}%
\bibitem [{\citenamefont {Zhou}\ \emph {et~al.}(2025)\citenamefont {Zhou},
  \citenamefont {Suresh}, \citenamefont {Robicheaux},\ and\ \citenamefont
  {Hung}}]{Zhou2025}%
  \BibitemOpen
  \bibfield  {author} {\bibinfo {author} {\bibfnamefont {X.}~\bibnamefont
  {Zhou}}, \bibinfo {author} {\bibfnamefont {D.~A.}\ \bibnamefont {Suresh}},
  \bibinfo {author} {\bibfnamefont {F.}~\bibnamefont {Robicheaux}}, \ and\
  \bibinfo {author} {\bibfnamefont {C.-L.}\ \bibnamefont {Hung}},\ }\bibfield
  {title} {\enquote {\bibinfo {title} {{Selective Collective Emission from a
  Dense Atomic Ensemble Coupled to a Nanophotonic Resonator}},}\ }\href
  {\doibase 10.1103/cdd5-r7h4} {\bibfield  {journal} {\bibinfo  {journal}
  {Phys. Rev. Lett.}\ }\textbf {\bibinfo {volume} {135}},\ \bibinfo {pages}
  {113601} (\bibinfo {year} {2025})}\BibitemShut {NoStop}%
\bibitem [{\citenamefont {Ozawa}\ \emph {et~al.}(2019)\citenamefont {Ozawa},
  \citenamefont {Price}, \citenamefont {Amo}, \citenamefont {Goldman},
  \citenamefont {Hafezi}, \citenamefont {Lu}, \citenamefont {Rechtsman},
  \citenamefont {Schuster}, \citenamefont {Simon}, \citenamefont {Zilberberg}
  \emph {et~al.}}]{ozawa2019topological}%
  \BibitemOpen
  \bibfield  {author} {\bibinfo {author} {\bibfnamefont {T.}~\bibnamefont
  {Ozawa}}, \bibinfo {author} {\bibfnamefont {H.~M.}\ \bibnamefont {Price}},
  \bibinfo {author} {\bibfnamefont {A.}~\bibnamefont {Amo}}, \bibinfo {author}
  {\bibfnamefont {N.}~\bibnamefont {Goldman}}, \bibinfo {author} {\bibfnamefont
  {M.}~\bibnamefont {Hafezi}}, \bibinfo {author} {\bibfnamefont
  {L.}~\bibnamefont {Lu}}, \bibinfo {author} {\bibfnamefont {M.~C.}\
  \bibnamefont {Rechtsman}}, \bibinfo {author} {\bibfnamefont {D.}~\bibnamefont
  {Schuster}}, \bibinfo {author} {\bibfnamefont {J.}~\bibnamefont {Simon}},
  \bibinfo {author} {\bibfnamefont {O.}~\bibnamefont {Zilberberg}},  \emph
  {et~al.},\ }\bibfield  {title} {\enquote {\bibinfo {title} {Topological
  photonics},}\ }\href {\doibase 10.1103/RevModPhys.91.015006} {\bibfield
  {journal} {\bibinfo  {journal} {Rev. Mod. Phys}\ }\textbf {\bibinfo {volume}
  {91}},\ \bibinfo {pages} {015006} (\bibinfo {year} {2019})}\BibitemShut
  {NoStop}%
\bibitem [{\citenamefont {Hafezi}\ \emph {et~al.}(2013)\citenamefont {Hafezi},
  \citenamefont {Mittal}, \citenamefont {Fan}, \citenamefont {Migdall},\ and\
  \citenamefont {Taylor}}]{hafezi2013imaging}%
  \BibitemOpen
  \bibfield  {author} {\bibinfo {author} {\bibfnamefont {M.}~\bibnamefont
  {Hafezi}}, \bibinfo {author} {\bibfnamefont {S.}~\bibnamefont {Mittal}},
  \bibinfo {author} {\bibfnamefont {J.}~\bibnamefont {Fan}}, \bibinfo {author}
  {\bibfnamefont {A.}~\bibnamefont {Migdall}}, \ and\ \bibinfo {author}
  {\bibfnamefont {J.}~\bibnamefont {Taylor}},\ }\bibfield  {title} {\enquote
  {\bibinfo {title} {Imaging topological edge states in silicon photonics},}\
  }\href {\doibase 10.1038/nphoton.2013.274} {\bibfield  {journal} {\bibinfo
  {journal} {Nat. Photon.}\ }\textbf {\bibinfo {volume} {7}},\ \bibinfo {pages}
  {1001} (\bibinfo {year} {2013})}\BibitemShut {NoStop}%
\bibitem [{\citenamefont {Hodaei}\ \emph {et~al.}(2017)\citenamefont {Hodaei},
  \citenamefont {Hassan}, \citenamefont {Wittek}, \citenamefont
  {Garcia-Gracia}, \citenamefont {El-Ganainy}, \citenamefont
  {Christodoulides},\ and\ \citenamefont {Khajavikhan}}]{hodaei2017enhanced}%
  \BibitemOpen
  \bibfield  {author} {\bibinfo {author} {\bibfnamefont {H.}~\bibnamefont
  {Hodaei}}, \bibinfo {author} {\bibfnamefont {A.~U.}\ \bibnamefont {Hassan}},
  \bibinfo {author} {\bibfnamefont {S.}~\bibnamefont {Wittek}}, \bibinfo
  {author} {\bibfnamefont {H.}~\bibnamefont {Garcia-Gracia}}, \bibinfo {author}
  {\bibfnamefont {R.}~\bibnamefont {El-Ganainy}}, \bibinfo {author}
  {\bibfnamefont {D.~N.}\ \bibnamefont {Christodoulides}}, \ and\ \bibinfo
  {author} {\bibfnamefont {M.}~\bibnamefont {Khajavikhan}},\ }\bibfield
  {title} {\enquote {\bibinfo {title} {Enhanced sensitivity at higher-order
  exceptional points},}\ }\href {\doibase 10.1038/nature23280} {\bibfield
  {journal} {\bibinfo  {journal} {Nature}\ }\textbf {\bibinfo {volume} {548}},\
  \bibinfo {pages} {187} (\bibinfo {year} {2017})}\BibitemShut {NoStop}%
\bibitem [{\citenamefont {Zhang}\ \emph
  {et~al.}(2019{\natexlab{b}})\citenamefont {Zhang}, \citenamefont {Wang},
  \citenamefont {Hu}, \citenamefont {Shams-Ansari}, \citenamefont {Ren},
  \citenamefont {Fan},\ and\ \citenamefont
  {Lon{\v{c}}ar}}]{zhang2019electronically}%
  \BibitemOpen
  \bibfield  {author} {\bibinfo {author} {\bibfnamefont {M.}~\bibnamefont
  {Zhang}}, \bibinfo {author} {\bibfnamefont {C.}~\bibnamefont {Wang}},
  \bibinfo {author} {\bibfnamefont {Y.}~\bibnamefont {Hu}}, \bibinfo {author}
  {\bibfnamefont {A.}~\bibnamefont {Shams-Ansari}}, \bibinfo {author}
  {\bibfnamefont {T.}~\bibnamefont {Ren}}, \bibinfo {author} {\bibfnamefont
  {S.}~\bibnamefont {Fan}}, \ and\ \bibinfo {author} {\bibfnamefont
  {M.}~\bibnamefont {Lon{\v{c}}ar}},\ }\bibfield  {title} {\enquote {\bibinfo
  {title} {Electronically programmable photonic molecule},}\ }\href {\doibase
  10.1038/s41566-018-0317-y} {\bibfield  {journal} {\bibinfo  {journal} {Nat.
  Photon.}\ }\textbf {\bibinfo {volume} {13}},\ \bibinfo {pages} {36} (\bibinfo
  {year} {2019}{\natexlab{b}})}\BibitemShut {NoStop}%
\bibitem [{\citenamefont {Vlasov}\ \emph {et~al.}(2008)\citenamefont {Vlasov},
  \citenamefont {Green},\ and\ \citenamefont {Xia}}]{vlasov2008high}%
  \BibitemOpen
  \bibfield  {author} {\bibinfo {author} {\bibfnamefont {Y.}~\bibnamefont
  {Vlasov}}, \bibinfo {author} {\bibfnamefont {W.~M.}\ \bibnamefont {Green}}, \
  and\ \bibinfo {author} {\bibfnamefont {F.}~\bibnamefont {Xia}},\ }\bibfield
  {title} {\enquote {\bibinfo {title} {High-throughput silicon nanophotonic
  wavelength-insensitive switch for on-chip optical networks},}\ }\href
  {\doibase 10.1038/nphoton.2008.31} {\bibfield  {journal} {\bibinfo  {journal}
  {Nat. Photon.}\ }\textbf {\bibinfo {volume} {2}},\ \bibinfo {pages} {242}
  (\bibinfo {year} {2008})}\BibitemShut {NoStop}%
\bibitem [{\citenamefont {Zhang}\ \emph {et~al.}(2016)\citenamefont {Zhang},
  \citenamefont {Zhu}, \citenamefont {Zou},\ and\ \citenamefont
  {Tang}}]{zhang2016optomagnonic}%
  \BibitemOpen
  \bibfield  {author} {\bibinfo {author} {\bibfnamefont {X.}~\bibnamefont
  {Zhang}}, \bibinfo {author} {\bibfnamefont {N.}~\bibnamefont {Zhu}}, \bibinfo
  {author} {\bibfnamefont {C.-L.}\ \bibnamefont {Zou}}, \ and\ \bibinfo
  {author} {\bibfnamefont {H.~X.}\ \bibnamefont {Tang}},\ }\bibfield  {title}
  {\enquote {\bibinfo {title} {Optomagnonic whispering gallery
  microresonators},}\ }\href {\doibase 10.1103/PhysRevLett.117.123605}
  {\bibfield  {journal} {\bibinfo  {journal} {Phys. Rev. Lett.}\ }\textbf
  {\bibinfo {volume} {117}},\ \bibinfo {pages} {123605} (\bibinfo {year}
  {2016})}\BibitemShut {NoStop}%
\bibitem [{\citenamefont {Bi}\ \emph {et~al.}(2011)\citenamefont {Bi},
  \citenamefont {Hu}, \citenamefont {Jiang}, \citenamefont {Kim}, \citenamefont
  {Dionne}, \citenamefont {Kimerling},\ and\ \citenamefont
  {Ross}}]{bi2011chip}%
  \BibitemOpen
  \bibfield  {author} {\bibinfo {author} {\bibfnamefont {L.}~\bibnamefont
  {Bi}}, \bibinfo {author} {\bibfnamefont {J.}~\bibnamefont {Hu}}, \bibinfo
  {author} {\bibfnamefont {P.}~\bibnamefont {Jiang}}, \bibinfo {author}
  {\bibfnamefont {D.~H.}\ \bibnamefont {Kim}}, \bibinfo {author} {\bibfnamefont
  {G.~F.}\ \bibnamefont {Dionne}}, \bibinfo {author} {\bibfnamefont {L.~C.}\
  \bibnamefont {Kimerling}}, \ and\ \bibinfo {author} {\bibfnamefont
  {C.}~\bibnamefont {Ross}},\ }\bibfield  {title} {\enquote {\bibinfo {title}
  {On-chip optical isolation in monolithically integrated non-reciprocal
  optical resonators},}\ }\href {\doibase 10.1038/nphoton.2011.270} {\bibfield
  {journal} {\bibinfo  {journal} {Nat. Photon.}\ }\textbf {\bibinfo {volume}
  {5}},\ \bibinfo {pages} {758} (\bibinfo {year} {2011})}\BibitemShut {NoStop}%
\bibitem [{\citenamefont {Liu}\ \emph {et~al.}(2011)\citenamefont {Liu},
  \citenamefont {Yin}, \citenamefont {Ulin-Avila}, \citenamefont {Geng},
  \citenamefont {Zentgraf}, \citenamefont {Ju}, \citenamefont {Wang},\ and\
  \citenamefont {Zhang}}]{liu2011graphene}%
  \BibitemOpen
  \bibfield  {author} {\bibinfo {author} {\bibfnamefont {M.}~\bibnamefont
  {Liu}}, \bibinfo {author} {\bibfnamefont {X.}~\bibnamefont {Yin}}, \bibinfo
  {author} {\bibfnamefont {E.}~\bibnamefont {Ulin-Avila}}, \bibinfo {author}
  {\bibfnamefont {B.}~\bibnamefont {Geng}}, \bibinfo {author} {\bibfnamefont
  {T.}~\bibnamefont {Zentgraf}}, \bibinfo {author} {\bibfnamefont
  {L.}~\bibnamefont {Ju}}, \bibinfo {author} {\bibfnamefont {F.}~\bibnamefont
  {Wang}}, \ and\ \bibinfo {author} {\bibfnamefont {X.}~\bibnamefont {Zhang}},\
  }\bibfield  {title} {\enquote {\bibinfo {title} {A graphene-based broadband
  optical modulator},}\ }\href {\doibase 10.1038/nature10067} {\bibfield
  {journal} {\bibinfo  {journal} {Nature}\ }\textbf {\bibinfo {volume} {474}},\
  \bibinfo {pages} {64} (\bibinfo {year} {2011})}\BibitemShut {NoStop}%
\bibitem [{\citenamefont {Phare}\ \emph {et~al.}(2015)\citenamefont {Phare},
  \citenamefont {Daniel~Lee}, \citenamefont {Cardenas},\ and\ \citenamefont
  {Lipson}}]{phare2015graphene}%
  \BibitemOpen
  \bibfield  {author} {\bibinfo {author} {\bibfnamefont {C.~T.}\ \bibnamefont
  {Phare}}, \bibinfo {author} {\bibfnamefont {Y.-H.}\ \bibnamefont
  {Daniel~Lee}}, \bibinfo {author} {\bibfnamefont {J.}~\bibnamefont
  {Cardenas}}, \ and\ \bibinfo {author} {\bibfnamefont {M.}~\bibnamefont
  {Lipson}},\ }\bibfield  {title} {\enquote {\bibinfo {title} {Graphene
  electro-optic modulator with 30 ghz bandwidth},}\ }\href {\doibase
  10.1038/nphoton.2015.122} {\bibfield  {journal} {\bibinfo  {journal} {Nat.
  Photon.}\ }\textbf {\bibinfo {volume} {9}},\ \bibinfo {pages} {511} (\bibinfo
  {year} {2015})}\BibitemShut {NoStop}%
\bibitem [{\citenamefont {Poot}\ and\ \citenamefont {Tang}(2014)}]{Poot2014}%
  \BibitemOpen
  \bibfield  {author} {\bibinfo {author} {\bibfnamefont {M.}~\bibnamefont
  {Poot}}\ and\ \bibinfo {author} {\bibfnamefont {H.~X.}\ \bibnamefont
  {Tang}},\ }\bibfield  {title} {\enquote {\bibinfo {title} {{Broadband
  nanoelectromechanical phase shifting of light on a chip}},}\ }\href {\doibase
  10.1063/1.4864257} {\bibfield  {journal} {\bibinfo  {journal} {Appl. Phys.
  Lett.}\ }\textbf {\bibinfo {volume} {104}},\ \bibinfo {pages} {061101}
  (\bibinfo {year} {2014})}\BibitemShut {NoStop}%
\bibitem [{\citenamefont {Seok}\ \emph {et~al.}(2016)\citenamefont {Seok},
  \citenamefont {Quack}, \citenamefont {Han}, \citenamefont {Muller},\ and\
  \citenamefont {Wu}}]{seok2016large}%
  \BibitemOpen
  \bibfield  {author} {\bibinfo {author} {\bibfnamefont {T.~J.}\ \bibnamefont
  {Seok}}, \bibinfo {author} {\bibfnamefont {N.}~\bibnamefont {Quack}},
  \bibinfo {author} {\bibfnamefont {S.}~\bibnamefont {Han}}, \bibinfo {author}
  {\bibfnamefont {R.~S.}\ \bibnamefont {Muller}}, \ and\ \bibinfo {author}
  {\bibfnamefont {M.~C.}\ \bibnamefont {Wu}},\ }\bibfield  {title} {\enquote
  {\bibinfo {title} {Large-scale broadband digital silicon photonic switches
  with vertical adiabatic couplers},}\ }\href {\doibase
  10.1364/OPTICA.3.000064} {\bibfield  {journal} {\bibinfo  {journal} {Optica}\
  }\textbf {\bibinfo {volume} {3}},\ \bibinfo {pages} {64} (\bibinfo {year}
  {2016})}\BibitemShut {NoStop}%
\bibitem [{\citenamefont {R{\'\i}os}\ \emph {et~al.}(2015)\citenamefont
  {R{\'\i}os}, \citenamefont {Stegmaier}, \citenamefont {Hosseini},
  \citenamefont {Wang}, \citenamefont {Scherer}, \citenamefont {Wright},
  \citenamefont {Bhaskaran},\ and\ \citenamefont
  {Pernice}}]{rios2015integrated}%
  \BibitemOpen
  \bibfield  {author} {\bibinfo {author} {\bibfnamefont {C.}~\bibnamefont
  {R{\'\i}os}}, \bibinfo {author} {\bibfnamefont {M.}~\bibnamefont
  {Stegmaier}}, \bibinfo {author} {\bibfnamefont {P.}~\bibnamefont {Hosseini}},
  \bibinfo {author} {\bibfnamefont {D.}~\bibnamefont {Wang}}, \bibinfo {author}
  {\bibfnamefont {T.}~\bibnamefont {Scherer}}, \bibinfo {author} {\bibfnamefont
  {C.~D.}\ \bibnamefont {Wright}}, \bibinfo {author} {\bibfnamefont
  {H.}~\bibnamefont {Bhaskaran}}, \ and\ \bibinfo {author} {\bibfnamefont
  {W.~H.}\ \bibnamefont {Pernice}},\ }\bibfield  {title} {\enquote {\bibinfo
  {title} {Integrated all-photonic non-volatile multi-level memory},}\ }\href
  {\doibase 10.1038/nphoton.2015.182} {\bibfield  {journal} {\bibinfo
  {journal} {Nat. Photon.}\ }\textbf {\bibinfo {volume} {9}},\ \bibinfo {pages}
  {725} (\bibinfo {year} {2015})}\BibitemShut {NoStop}%
\bibitem [{\citenamefont {Zhang}\ \emph {et~al.}(2025)\citenamefont {Zhang},
  \citenamefont {Xu}, \citenamefont {Gong}, \citenamefont {Wang}, \citenamefont
  {Qi}, \citenamefont {Liu}, \citenamefont {Yang}, \citenamefont {Tian},
  \citenamefont {Wang}, \citenamefont {Zhang}, \citenamefont {Li},
  \citenamefont {Guo}, \citenamefont {Yan}, \citenamefont {Dong}, \citenamefont
  {Ren}, \citenamefont {Zhang}, \citenamefont {Zhang}, \citenamefont {Guo},
  \citenamefont {Che},\ and\ \citenamefont {Zou}}]{Zhang2025}%
  \BibitemOpen
  \bibfield  {author} {\bibinfo {author} {\bibfnamefont {J.-Z.}\ \bibnamefont
  {Zhang}}, \bibinfo {author} {\bibfnamefont {X.-B.}\ \bibnamefont {Xu}},
  \bibinfo {author} {\bibfnamefont {Y.}~\bibnamefont {Gong}}, \bibinfo {author}
  {\bibfnamefont {Z.-B.}\ \bibnamefont {Wang}}, \bibinfo {author}
  {\bibfnamefont {X.-Z.}\ \bibnamefont {Qi}}, \bibinfo {author} {\bibfnamefont
  {X.-J.}\ \bibnamefont {Liu}}, \bibinfo {author} {\bibfnamefont {Y.-H.}\
  \bibnamefont {Yang}}, \bibinfo {author} {\bibfnamefont {Z.-H.}\ \bibnamefont
  {Tian}}, \bibinfo {author} {\bibfnamefont {J.-Q.}\ \bibnamefont {Wang}},
  \bibinfo {author} {\bibfnamefont {Y.-L.}\ \bibnamefont {Zhang}}, \bibinfo
  {author} {\bibfnamefont {M.}~\bibnamefont {Li}}, \bibinfo {author}
  {\bibfnamefont {Y.}~\bibnamefont {Guo}}, \bibinfo {author} {\bibfnamefont
  {Y.}~\bibnamefont {Yan}}, \bibinfo {author} {\bibfnamefont {C.-H.}\
  \bibnamefont {Dong}}, \bibinfo {author} {\bibfnamefont {X.-F.}\ \bibnamefont
  {Ren}}, \bibinfo {author} {\bibfnamefont {Y.}~\bibnamefont {Zhang}}, \bibinfo
  {author} {\bibfnamefont {C.}~\bibnamefont {Zhang}}, \bibinfo {author}
  {\bibfnamefont {G.-C.}\ \bibnamefont {Guo}}, \bibinfo {author} {\bibfnamefont
  {Y.}~\bibnamefont {Che}}, \ and\ \bibinfo {author} {\bibfnamefont {C.-L.}\
  \bibnamefont {Zou}},\ }\bibfield  {title} {\enquote {\bibinfo {title}
  {{Optically-driven organic nano-step actuator for reconfigurable photonic
  circuits}},}\ }\href {\doibase 10.1038/s41467-025-63521-z} {\bibfield
  {journal} {\bibinfo  {journal} {Nat. Commun.}\ }\textbf {\bibinfo {volume}
  {16}},\ \bibinfo {pages} {8213} (\bibinfo {year} {2025})}\BibitemShut
  {NoStop}%
\bibitem [{\citenamefont {Yang}\ \emph {et~al.}(2024)\citenamefont {Yang},
  \citenamefont {Wang}, \citenamefont {Xu}, \citenamefont {Li}, \citenamefont
  {Zhang}, \citenamefont {Pan}, \citenamefont {Xiao}, \citenamefont {Wang},
  \citenamefont {Guo}, \citenamefont {Sun} \emph {et~al.}}]{yang2024proposal}%
  \BibitemOpen
  \bibfield  {author} {\bibinfo {author} {\bibfnamefont {Y.-H.}\ \bibnamefont
  {Yang}}, \bibinfo {author} {\bibfnamefont {J.-Q.}\ \bibnamefont {Wang}},
  \bibinfo {author} {\bibfnamefont {X.-B.}\ \bibnamefont {Xu}}, \bibinfo
  {author} {\bibfnamefont {M.}~\bibnamefont {Li}}, \bibinfo {author}
  {\bibfnamefont {Y.-L.}\ \bibnamefont {Zhang}}, \bibinfo {author}
  {\bibfnamefont {X.}~\bibnamefont {Pan}}, \bibinfo {author} {\bibfnamefont
  {L.}~\bibnamefont {Xiao}}, \bibinfo {author} {\bibfnamefont {W.}~\bibnamefont
  {Wang}}, \bibinfo {author} {\bibfnamefont {G.-C.}\ \bibnamefont {Guo}},
  \bibinfo {author} {\bibfnamefont {L.}~\bibnamefont {Sun}},  \emph {et~al.},\
  }\bibfield  {title} {\enquote {\bibinfo {title} {Proposal for brillouin
  microwave-to-optical conversion on a chip},}\ }\href {\doibase
  10.1364/OME.534817} {\bibfield  {journal} {\bibinfo  {journal} {Opt. Mater.
  Express}\ }\textbf {\bibinfo {volume} {14}},\ \bibinfo {pages} {2400}
  (\bibinfo {year} {2024})}\BibitemShut {NoStop}%
\bibitem [{\citenamefont {Yang}\ \emph {et~al.}(2025)\citenamefont {Yang},
  \citenamefont {Wang}, \citenamefont {Zhu}, \citenamefont {Zeng},
  \citenamefont {Li}, \citenamefont {Zhang}, \citenamefont {Lu}, \citenamefont
  {Zhang}, \citenamefont {Wang}, \citenamefont {Dong}, \citenamefont {Xu},
  \citenamefont {Guo}, \citenamefont {Sun},\ and\ \citenamefont
  {Zou}}]{yang2025multi}%
  \BibitemOpen
  \bibfield  {author} {\bibinfo {author} {\bibfnamefont {Y.-H.}\ \bibnamefont
  {Yang}}, \bibinfo {author} {\bibfnamefont {J.-Q.}\ \bibnamefont {Wang}},
  \bibinfo {author} {\bibfnamefont {Z.-X.}\ \bibnamefont {Zhu}}, \bibinfo
  {author} {\bibfnamefont {Y.}~\bibnamefont {Zeng}}, \bibinfo {author}
  {\bibfnamefont {M.}~\bibnamefont {Li}}, \bibinfo {author} {\bibfnamefont
  {Y.-L.}\ \bibnamefont {Zhang}}, \bibinfo {author} {\bibfnamefont
  {J.}~\bibnamefont {Lu}}, \bibinfo {author} {\bibfnamefont {Q.}~\bibnamefont
  {Zhang}}, \bibinfo {author} {\bibfnamefont {W.}~\bibnamefont {Wang}},
  \bibinfo {author} {\bibfnamefont {C.-H.}\ \bibnamefont {Dong}}, \bibinfo
  {author} {\bibfnamefont {X.-B.}\ \bibnamefont {Xu}}, \bibinfo {author}
  {\bibfnamefont {G.-C.}\ \bibnamefont {Guo}}, \bibinfo {author} {\bibfnamefont
  {L.}~\bibnamefont {Sun}}, \ and\ \bibinfo {author} {\bibfnamefont {C.-L.}\
  \bibnamefont {Zou}},\ }\bibfield  {title} {\enquote {\bibinfo {title}
  {Multi-channel microwave-to-optics conversion utilizing a hybrid
  photonic-phononic waveguide},}\ }\href {https://arxiv.org/abs/2509.10052}
  {\bibfield  {journal} {\bibinfo  {journal} {arXiv preprint: 2509.10052}\ }
  (\bibinfo {year} {2025})}\BibitemShut {NoStop}%
\bibitem [{\citenamefont {Guo}\ \emph {et~al.}(2016)\citenamefont {Guo},
  \citenamefont {Zou}, \citenamefont {Jung},\ and\ \citenamefont
  {Tang}}]{Guo2016}%
  \BibitemOpen
  \bibfield  {author} {\bibinfo {author} {\bibfnamefont {X.}~\bibnamefont
  {Guo}}, \bibinfo {author} {\bibfnamefont {C.-L.}\ \bibnamefont {Zou}},
  \bibinfo {author} {\bibfnamefont {H.}~\bibnamefont {Jung}}, \ and\ \bibinfo
  {author} {\bibfnamefont {H.~X.}\ \bibnamefont {Tang}},\ }\bibfield  {title}
  {\enquote {\bibinfo {title} {{On-Chip Strong Coupling and Efficient Frequency
  Conversion between Telecom and Visible Optical Modes}},}\ }\href {\doibase
  10.1103/PhysRevLett.117.123902} {\bibfield  {journal} {\bibinfo  {journal}
  {Physical Review Letters}\ }\textbf {\bibinfo {volume} {117}},\ \bibinfo
  {pages} {123902} (\bibinfo {year} {2016})}\BibitemShut {NoStop}%
\bibitem [{\citenamefont {Wang}\ \emph {et~al.}(2025)\citenamefont {Wang},
  \citenamefont {Yang}, \citenamefont {Zhu}, \citenamefont {Lu}, \citenamefont
  {Li}, \citenamefont {Pan}, \citenamefont {Ma}, \citenamefont {Xiao},
  \citenamefont {Zhang}, \citenamefont {Wang}, \citenamefont {Dong},
  \citenamefont {Xu}, \citenamefont {Guo}, \citenamefont {Sun},\ and\
  \citenamefont {Zou}}]{Wang2025}%
  \BibitemOpen
  \bibfield  {author} {\bibinfo {author} {\bibfnamefont {J.-Q.}\ \bibnamefont
  {Wang}}, \bibinfo {author} {\bibfnamefont {Y.-H.}\ \bibnamefont {Yang}},
  \bibinfo {author} {\bibfnamefont {Z.-X.}\ \bibnamefont {Zhu}}, \bibinfo
  {author} {\bibfnamefont {J.-J.}\ \bibnamefont {Lu}}, \bibinfo {author}
  {\bibfnamefont {M.}~\bibnamefont {Li}}, \bibinfo {author} {\bibfnamefont
  {X.}~\bibnamefont {Pan}}, \bibinfo {author} {\bibfnamefont {C.}~\bibnamefont
  {Ma}}, \bibinfo {author} {\bibfnamefont {L.}~\bibnamefont {Xiao}}, \bibinfo
  {author} {\bibfnamefont {B.}~\bibnamefont {Zhang}}, \bibinfo {author}
  {\bibfnamefont {W.}~\bibnamefont {Wang}}, \bibinfo {author} {\bibfnamefont
  {C.-H.}\ \bibnamefont {Dong}}, \bibinfo {author} {\bibfnamefont {X.-B.}\
  \bibnamefont {Xu}}, \bibinfo {author} {\bibfnamefont {G.-C.}\ \bibnamefont
  {Guo}}, \bibinfo {author} {\bibfnamefont {L.}~\bibnamefont {Sun}}, \ and\
  \bibinfo {author} {\bibfnamefont {C.-L.}\ \bibnamefont {Zou}},\ }\bibfield
  {title} {\enquote {\bibinfo {title} {{Compact and high-resolution
  spectrometer via Brillouin integrated circuits}},}\ }\href
  {http://arxiv.org/abs/2511.04444} {\bibfield  {journal} {\bibinfo  {journal}
  {arXiv preprint: 2511.04444}\ } (\bibinfo {year} {2025})}\BibitemShut
  {NoStop}%
\bibitem [{\citenamefont {Xu}\ \emph {et~al.}(2019)\citenamefont {Xu},
  \citenamefont {Shi}, \citenamefont {Guo}, \citenamefont {Dong},\ and\
  \citenamefont {Zou}}]{Xu2019}%
  \BibitemOpen
  \bibfield  {author} {\bibinfo {author} {\bibfnamefont {X.-B.}\ \bibnamefont
  {Xu}}, \bibinfo {author} {\bibfnamefont {L.}~\bibnamefont {Shi}}, \bibinfo
  {author} {\bibfnamefont {G.-C.}\ \bibnamefont {Guo}}, \bibinfo {author}
  {\bibfnamefont {C.-H.}\ \bibnamefont {Dong}}, \ and\ \bibinfo {author}
  {\bibfnamefont {C.-L.}\ \bibnamefont {Zou}},\ }\bibfield  {title} {\enquote
  {\bibinfo {title} {{Mobius microring resonator}},}\ }\href
  {\doibase 10.1063/1.5082675} {\bibfield  {journal} {\bibinfo  {journal}
  {Appl. Phys. Lett.}\ }\textbf {\bibinfo {volume} {114}},\ \bibinfo {pages}
  {101106} (\bibinfo {year} {2019})}\BibitemShut {NoStop}%
\bibitem [{\citenamefont {Lin}\ \emph {et~al.}(2022)\citenamefont {Lin},
  \citenamefont {Zhou}, \citenamefont {Liu}, \citenamefont {Shu}, \citenamefont
  {Zou}, \citenamefont {Dong}, \citenamefont {Wei}, \citenamefont {Dong},
  \citenamefont {Zhang}, \citenamefont {Yao},\ and\ \citenamefont
  {Zhao}}]{Lin2022}%
  \BibitemOpen
  \bibfield  {author} {\bibinfo {author} {\bibfnamefont {X.}~\bibnamefont
  {Lin}}, \bibinfo {author} {\bibfnamefont {W.}~\bibnamefont {Zhou}}, \bibinfo
  {author} {\bibfnamefont {Y.}~\bibnamefont {Liu}}, \bibinfo {author}
  {\bibfnamefont {F.}~\bibnamefont {Shu}}, \bibinfo {author} {\bibfnamefont
  {C.}~\bibnamefont {Zou}}, \bibinfo {author} {\bibfnamefont {C.}~\bibnamefont
  {Dong}}, \bibinfo {author} {\bibfnamefont {C.}~\bibnamefont {Wei}}, \bibinfo
  {author} {\bibfnamefont {H.}~\bibnamefont {Dong}}, \bibinfo {author}
  {\bibfnamefont {C.}~\bibnamefont {Zhang}}, \bibinfo {author} {\bibfnamefont
  {J.}~\bibnamefont {Yao}}, \ and\ \bibinfo {author} {\bibfnamefont {Y.~S.}\
  \bibnamefont {Zhao}},\ }\bibfield  {title} {\enquote {\bibinfo {title}
  {{3D-Printed Mobius Microring Lasers: Topology Engineering in
  Photonic Microstructures}},}\ }\href {\doibase 10.1002/smll.202202812}
  {\bibfield  {journal} {\bibinfo  {journal} {Small}\ }\textbf {\bibinfo
  {volume} {18}},\ \bibinfo {pages} {202202812} (\bibinfo {year}
  {2022})}\BibitemShut {NoStop}%
\bibitem [{\citenamefont {Chen}\ \emph {et~al.}(2023)\citenamefont {Chen},
  \citenamefont {Hou}, \citenamefont {Zhao}, \citenamefont {Chen},\ and\
  \citenamefont {Wan}}]{Chen2023}%
  \BibitemOpen
  \bibfield  {author} {\bibinfo {author} {\bibfnamefont {Y.}~\bibnamefont
  {Chen}}, \bibinfo {author} {\bibfnamefont {J.}~\bibnamefont {Hou}}, \bibinfo
  {author} {\bibfnamefont {G.}~\bibnamefont {Zhao}}, \bibinfo {author}
  {\bibfnamefont {X.}~\bibnamefont {Chen}}, \ and\ \bibinfo {author}
  {\bibfnamefont {W.}~\bibnamefont {Wan}},\ }\bibfield  {title} {\enquote
  {\bibinfo {title} {{Topological resonances in a Mobius ring
  resonator}},}\ }\href {\doibase 10.1038/s42005-023-01205-0} {\bibfield
  {journal} {\bibinfo  {journal} {Commun. Phys.}\ }\textbf {\bibinfo {volume}
  {6}},\ \bibinfo {pages} {84} (\bibinfo {year} {2023})}\BibitemShut {NoStop}%
\bibitem [{\citenamefont {Yang}\ \emph {et~al.}(2023)\citenamefont {Yang},
  \citenamefont {Wang}, \citenamefont {Zhu}, \citenamefont {Xu}, \citenamefont
  {Zhang}, \citenamefont {Lu}, \citenamefont {Zeng}, \citenamefont {Dong},
  \citenamefont {Sun}, \citenamefont {Guo},\ and\ \citenamefont
  {Zou}}]{yang_stimulated_2023}%
  \BibitemOpen
  \bibfield  {author} {\bibinfo {author} {\bibfnamefont {Y.-H.}\ \bibnamefont
  {Yang}}, \bibinfo {author} {\bibfnamefont {J.-Q.}\ \bibnamefont {Wang}},
  \bibinfo {author} {\bibfnamefont {Z.-X.}\ \bibnamefont {Zhu}}, \bibinfo
  {author} {\bibfnamefont {X.-B.}\ \bibnamefont {Xu}}, \bibinfo {author}
  {\bibfnamefont {Q.}~\bibnamefont {Zhang}}, \bibinfo {author} {\bibfnamefont
  {J.}~\bibnamefont {Lu}}, \bibinfo {author} {\bibfnamefont {Y.}~\bibnamefont
  {Zeng}}, \bibinfo {author} {\bibfnamefont {C.-H.}\ \bibnamefont {Dong}},
  \bibinfo {author} {\bibfnamefont {L.}~\bibnamefont {Sun}}, \bibinfo {author}
  {\bibfnamefont {G.-C.}\ \bibnamefont {Guo}}, \ and\ \bibinfo {author}
  {\bibfnamefont {C.-L.}\ \bibnamefont {Zou}},\ }\bibfield  {title}
  {{\enquote {\bibinfo {title} {Stimulated Brillouin
  interaction between guided phonons and photons in a lithium niobate
  waveguide},}\ }}\href {\doibase 10.1007/s11433-023-2272-y} {\bibfield
  {journal} {\bibinfo  {journal} {Sci. China Phys. Mech.}\ }\textbf {\bibinfo
  {volume} {67}},\ \bibinfo {pages} {214221} (\bibinfo {year}
  {2023})}\BibitemShut {NoStop}%
\end{thebibliography}
\end{document}